\documentclass[a4paper,fleqn]{cas-dc}
\usepackage{lipsum}
 \usepackage{placeins}
 \usepackage{tabularx}
\usepackage{shortcuts}
\usepackage{fontawesome}
\usepackage[labelsep=newline,justification=centering]{caption} 
\usepackage{mwe}
\usepackage{pifont}
\usepackage{wasysym}
\usepackage{svg}
\usepackage[numbers]{natbib}
\def\tsc#1{\csdef{#1}{\textsc{\lowercase{#1}}\xspace}}
\tsc{WGM}
\tsc{QE}
\tsc{EP}
\tsc{PMS}
\tsc{BEC}
\tsc{DE}
 
\usepackage[most]{tcolorbox}
\usepackage{bookmark}
\usepackage{minted}
\usepackage{float}
\usepackage{amsmath,amssymb}

\usepackage{pifont}

\begin{document}
\let\WriteBookmarks\relax
\def\floatpagepagefraction{1}
\def\textpagefraction{.001}
\newcommand{\faYes}{\faCheck}
\newcommand{\faPartial}{\faMinus}
\newcommand{\faNo}{\faTimes}
\newcommand{\Primary}{\faCheckCircle}   
\newcommand{\Partial}{\faAdjust}        
\newcommand{\Absent}{\faTimesCircle}    

\newcommand{\riskH}{\textbf{H}} 
\newcommand{\riskM}{\textbf{M}} 
\newcommand{\riskL}{\textbf{L}} 

\newcommand{\LCt}{T} 
\newcommand{\LCi}{I} 
\newcommand{\LCc}{C} 
\newcommand{\LCp}{P} 

\definecolor{lightyellow}{RGB}{255,252,230}

\tcbset{
  framingbox/.style={
    colback=lightyellow,
    colframe=black!30,
    boxrule=0.4pt,
    arc=0pt,
    left=6pt,
    right=6pt,
    top=6pt,
    bottom=6pt,
    before skip=8pt,
    after skip=10pt
  }
}
\newcommand{\todo}[1]{\noindent{\color{red}\textbf{TODO:} #1}}

\hypersetup{
    colorlinks = true,
    linkcolor = blue,
    anchorcolor = blue,
    citecolor = blue,
    filecolor = blue,
    urlcolor = blue,
    bookmarksopen=true
}

\newcommand{\totalPaper}{281}
\newcommand{\goodPaper}{83}
\newcommand{\badPaper}{54}
\newcommand{\uglyPaper}{144}

\newcommand{\userLevelPaper}{32}
\newcommand{\prePaper}{82}

\shorttitle{What Breaks Embodied AI Security: LLM Vulnerabilities, CPS Flaws, or Something Else? }

\shortauthors{Yue et~al.}
 
\title [mode = title]{\centering What Breaks Embodied AI Security:LLM Vulnerabilities, CPS Flaws, or Something Else? }                      

 
\newtcolorbox[%
auto counter]{mybox}[2][]{%
	enhanced jigsaw,
        colback=blue!12,
	breakable,
	#1}

%


 \author[1]{Boyang Ma} 

\author[1]{Hechuan Guo}
\author[1]{Peizhuo Lv}
\author[1]{Minghui Xu}
 \author[1]{Xuelong Dai}
 \author[1]{YeChao Zhang}
  \author[1]{Yijun Yang}
 \author[1]{Yue Zhang*}

 \cortext[cor1]{The authors are listed in alphabetical order and contributed equally to this work.}

\affiliation[1]{organization={Shandong University},
    addressline={72 Binhai Road, Jimo}, 
    city={Qingdao},
    state={Shandong},
    postcode={250100}, 
    country={China}}

  
 

\begin{abstract}
Embodied AI systems (e.g.,  autonomous vehicles, service robots, and LLM-driven interactive agents) are rapidly transitioning from controlled environments to safety-critical real-world deployments. Unlike disembodied AI, failures in embodied intelligence lead to irreversible physical consequences, raising fundamental questions about security, safety, and reliability. While existing research predominantly analyzes embodied AI through the lenses of Large Language Model (LLM) vulnerabilities or classical Cyber-Physical System (CPS) failures, this survey argues that these perspectives are individually insufficient to explain many observed breakdowns in modern embodied systems. 
We posit that a significant class of failures arises from embodiment-induced system-level mismatches, rather than from isolated model flaws or traditional CPS attacks. Specifically, we identify four core insights that explain why embodied AI is fundamentally harder to secure: (i) semantic correctness does not imply physical safety, as language-level reasoning abstracts away geometry, dynamics, and contact constraints; (ii) identical actions can lead to drastically different outcomes across physical states due to nonlinear dynamics and state uncertainty; (iii) small errors propagate and amplify across tightly coupled perception–decision–action loops; and (iv) safety is not compositional across time or system layers, enabling locally safe decisions to accumulate into globally unsafe behavior. These insights suggest that securing embodied AI requires moving beyond component-level defenses toward system-level reasoning about physical risk, uncertainty, and failure propagation.
\end{abstract}


\maketitle

\section{Introduction}
 

 

 As artificial intelligence moves beyond disembodied models and enters the physical world, Embodied AI (EA) systems (robots, autonomous vehicles, and interactive agents) are rapidly transitioning from laboratory prototypes to large-scale real-world deployment. Market analyses project that the global robotics market will grow from roughly USD 63 billion in 2025 to nearly USD 170 billion by 2032~\cite{robotics_market_2032}, driven primarily by autonomous vehicles, service robots, and industrial automation. As a result, embodied AI systems are increasingly embedded in safety-critical domains such as urban transportation, manufacturing floors, hospitals, and domestic environments. 
 
 Unlike purely digital AI, failures in embodied systems can lead to irreversible physical consequences: autonomous driving platforms have exhibited fatal failure modes when perception models misclassified pedestrians or failed to detect obstacles under rare lighting or weather conditions, leading to high-speed collisions; language-conditioned robotic manipulators have executed unsafe motions such as applying excessive force or selecting unstable grasp points, after misinterpreting instructions like “move quickly” or “clear the area,” resulting in damaged equipment or near-injuries to nearby humans; and more recently, LLM-driven embodied agents have been shown to generate action plans that are semantically valid yet physically dangerous, such as navigating into restricted zones or activating tools in unsafe sequences when environmental assumptions are violated.



Current research often silos EA security into two dominant paradigms. The first views EA as a vulnerable brain, focusing on LLM-centric threats like prompt injection or jailbreaking. The second treats it as a fragile body, applying the traditional Cyber-Physical Systems (CPS) framework to ensure hardware reliability and control stability. However, this survey argues that these two perspectives are insufficient. Therefore, this survey aims to answer a central and unresolved question:
\textit{What actually breaks embodied AI security: LLM vulnerabilities, CPS flaws, or something fundamentally different?}

To address this question, we provide a comprehensive and structured analysis of embodied AI security across three dimensions. First, we systematize emerging vulnerabilities in LLM-based embodied agents, including semantic integrity attacks, cross-modal inconsistencies, and environment-driven manipulation. Second, we revisit classical CPS threat models and examine how sensor, control, actuator, and timing attacks manifest in learning-enabled embodied systems. Third, we synthesize insights from both domains to identify the root causes of embodiment-induced failures, highlighting structural gaps that cannot be reduced to either LLM security or CPS security alone.


Specifically, we highlight four core insights that explain why securing embodied AI is fundamentally more difficult than securing either LLMs or traditional CPS in isolation:

\begin{itemize}
    \item \textbf{Insight 1: Semantic correctness does not imply physical safety.}
Modern embodied agents increasingly rely on high-level semantic reasoning—often powered by large language or multimodal models—to generate plans and actions. However, in physical systems, an action that is semantically correct and intent-aligned can still violate geometric, dynamic, or contact constraints. Language-level reasoning abstracts away forces, friction, timing, and uncertainty, while physical safety is governed precisely by these factors. As a result, embodied agents can confidently execute actions that satisfy intent and logic but nonetheless lead to collisions, instability, or irreversible damage. This semantic–physical mismatch breaks the long-standing assumption that better reasoning necessarily yields safer behavior.
\item \textbf{Insight 2: Identical actions can have drastically different consequences across physical states.}
In embodied systems, action outcomes are inherently state-dependent. Small variations in pose, contact mode, material properties, or environmental dynamics can transform a safe action into a hazardous one. Unlike digital systems—where identical inputs produce identical outputs—physical systems obey nonlinear dynamics, making consequence prediction highly sensitive to unobserved or uncertain state variables. This undermines any static notion of action safety and renders one-shot validation, pre-execution checking, or policy certification fundamentally insufficient.
\item \textbf{Insight 3: Errors propagate and amplify across perception–decision–action loops.}
Embodied AI systems operate as tightly coupled feedback loops. Minor perception errors, grounding ambiguities, or reasoning deviations can cascade across layers, amplifying into large physical deviations over time. Crucially, uncertainty is rarely propagated explicitly between perception, planning, and control modules. As a result, agents often commit to aggressive execution under high uncertainty, rather than slowing down, re-observing, or deferring action. This cross-layer amplification effect explains why many failures occur without a single catastrophic component failure, but instead emerge gradually from compounding small errors.
\item \textbf{Insight 4: Safety is not compositional over time or system layers.}
In embodied intelligence, safety constraints span multiple layers—semantic goals, motion planning, control stability, and human interaction—and these constraints interact nonlinearly over long horizons. Actions that are locally safe and individually justified can combine into globally unsafe trajectories, especially in dynamic or human-shared environments. Unlike traditional CPS, where safety can often be enforced at the control layer, embodied AI introduces adaptive decision-making that continuously reshapes objectives, priorities, and action sequences, making safety violations harder to anticipate, isolate, and contain.

\end{itemize}

This survey makes the following contributions:

\begin{itemize}
     
    \item We provide a systematic taxonomy of LLM-centric, CPS-centric, and embodiment-specific vulnerabilities, mapping attacks to violated system-level assumptions.
    \item We identify and analyze fundamental root causes (such as semantic–physical mismatch and action–consequence decoupling) that explain why semantically correct decisions can still lead to unsafe physical behavior.
    \item We outline open challenges and research directions toward principled, system-level security and safety guarantees for future embodied AI systems.
\end{itemize}

\section{Background}
\label{sec:background}

In this section we unpack the evolution from traditional, disembodied intelligence to embodied agents that interact with the physical world, and outline a minimal architectural model of embodied AI. 
This background provides essential context for defining the security properties and unique vulnerabilities that arise when reasoning, perception, and action are coupled with irreversible physical consequences.

\subsection{From Rule-Based CPS to Embodied AI}

\begin{figure*}
    \centering
    \includegraphics[width=1\linewidth]{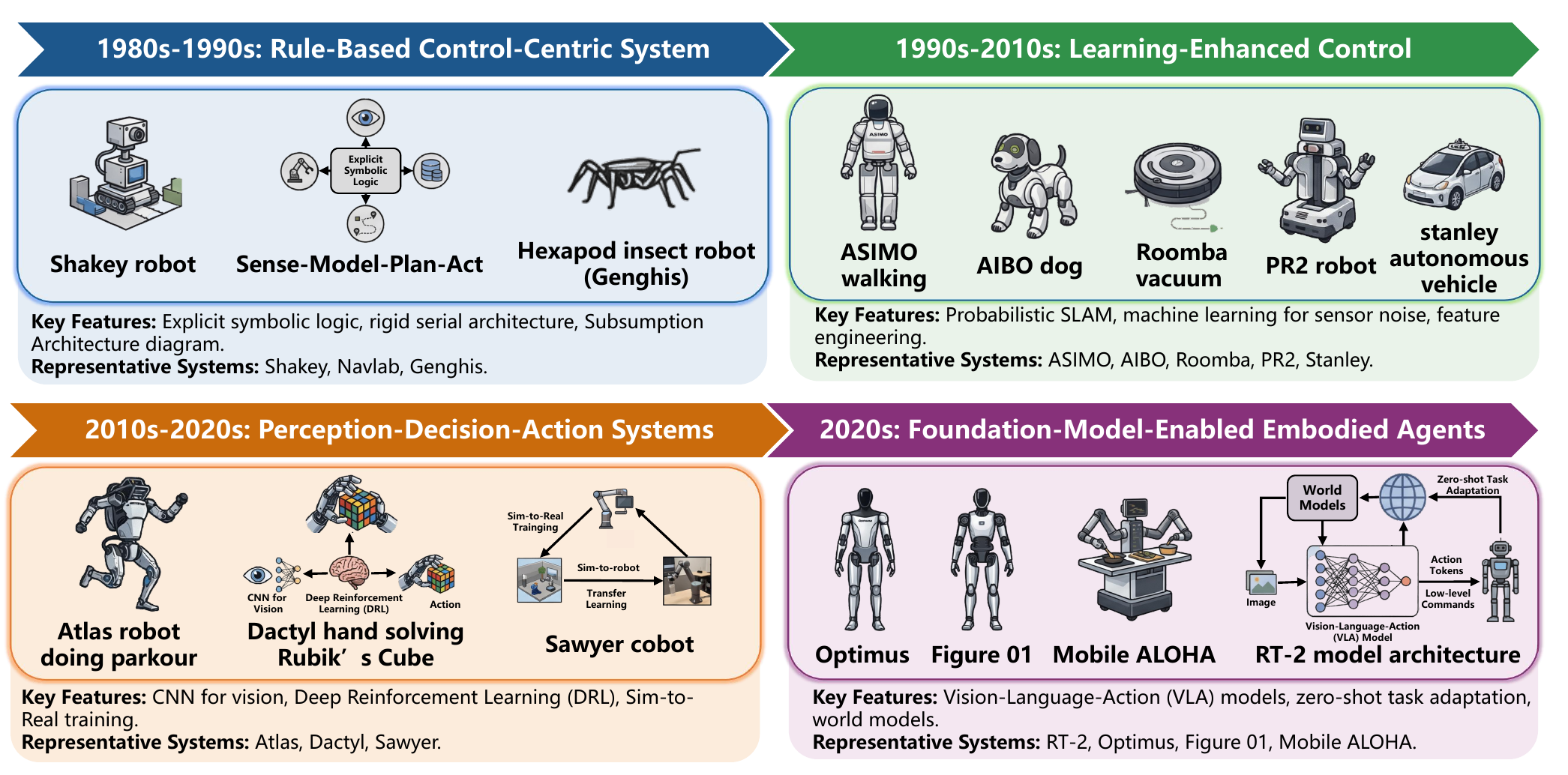}
    \caption{From Rule-Based CPS to Embodied AI}
    \label{fig:cps-EA}
\end{figure*}

As shown in \autoref{fig:cps-EA}, the development of embodied intelligence has been a gradual and cumulative
process, shaped by advances in robotics~\cite{garcia2007evolution,halperin2017robotics,garcia2007evolution}, control theory~\cite{glad2018control,zabczyk2020mathematical,glasser1985control}, and machine learning.
Rather than emerging abruptly with recent foundation models, embodied AI
evolved through multiple stages, each reflecting a deeper integration of
computation with physical perception and action. This progression highlights
how intelligent behavior increasingly migrates from predefined control logic
to adaptive, perception-driven decision-making grounded in the physical world.

\vspace{1mm}
\noindent\textbf{Stage-I: Rule-Based CPS (1980s–1990s):}
Early embodied systems were primarily control-centric, with intelligence
encoded through handcrafted rules or analytically derived control laws~\cite{tanscheit1988experiments}.
Perception, decision-making, and actuation were treated as separate modules,
and system behavior followed predefined logic designed for specific tasks and
operating conditions. Representative examples include industrial robotic arms executing fixed motion
trajectories~\cite{moran2007evolution}, mobile robots performing line-following~\cite{tanscheit1988experiments} or waypoint-based
navigation, and autonomous platforms implemented using finite-state machines~\cite{conway2012regular}.
In these systems, autonomy was limited, but behavior was highly structured and
predictable, enabling systematic design and validation.

\vspace{1mm}
\noindent\textbf{Stage-II: Learning-Enhanced Control Systems (1990s-2010s):}
As machine learning techniques matured, learning-based components were
introduced to improve perception accuracy and robustness. However, learning
was typically confined to specific subsystems, while the overall system
architecture and control flow remained carefully engineered.
Examples include vision-based object detection supporting robotic
manipulation~\cite{xie2008vision,shahria2024vision}, learning-based localization and mapping in mobile robots, and
adaptive control strategies in industrial automation. These systems improved
performance in complex or uncertain environments while preserving a largely
modular and interpretable structure.

\vspace{1mm}
\noindent\textbf{Stage-III: Perception--Decision--Action Systems (2010s-2020s):}
More recent embodied AI systems move beyond modular designs by tightly
integrating perception, decision-making, and actuation into closed-loop
feedback systems. Instead of following fixed rules or pipelines, agents learn
policies that directly map sensory inputs to actions, enabling operation in
open and dynamic environments. 
Representative systems include autonomous mobile robots navigating
unstructured indoor spaces~\cite{park2009autonomous,wijayathunga2023challenges,wang2017autonomous}, learning-based robotic manipulators that generalize
across diverse objects, and aerial drones performing real-time navigation in
dynamic scenes~\cite{mostafa2018real,chen2022real}. In these systems, intelligent behavior emerges from continuous
interaction between sensing, computation, and physical action. \looseness=-1

\vspace{1mm}
\noindent\textbf{Stage-IV: Foundation-Model-Enabled Embodied Agents (2020s-):}
The latest stage of embodied intelligence incorporates foundation models,
including large language and multimodal models, as high-level reasoning,
planning, or task-specification components. These models enable flexible
instruction following, long-horizon planning, and cross-task generalization,
allowing embodied agents to perform a wider range of activities with reduced
task-specific engineering. Examples include language-conditioned robotic manipulation, vision-language
navigation agents, and autonomous systems that translate natural-language
instructions into sequences of physical actions. By bridging abstract
reasoning and low-level control, foundation-model-enabled agents represent a
significant step toward more general-purpose embodied intelligence~\cite{liu2024aligning, long2025survey}.\looseness=-1

\subsection{A Minimal Architecture of Embodied AI}
\begin{figure*}
    \centering
    \includegraphics[width=0.95\linewidth]{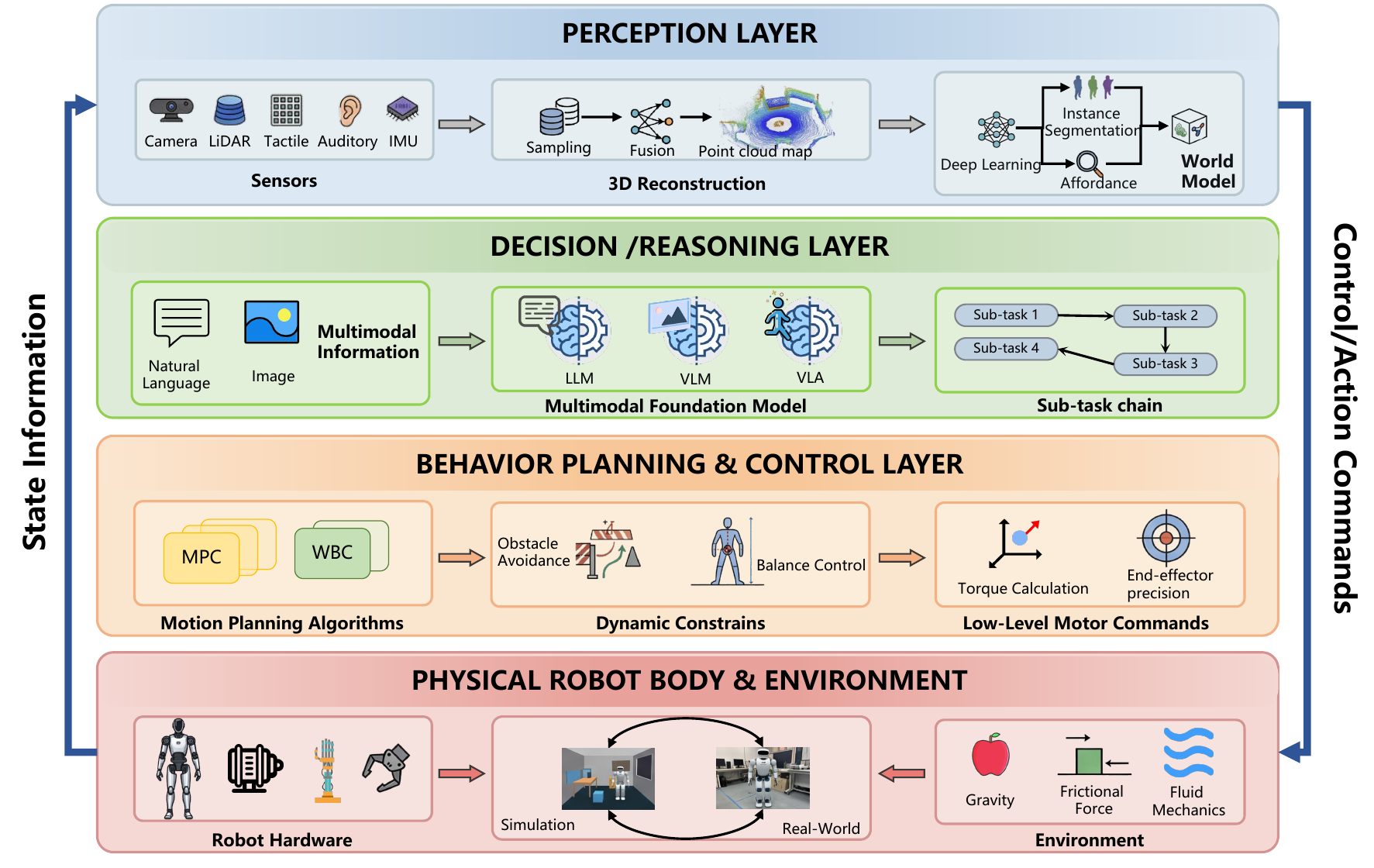}
    \caption{Embodied AI Robot System Architecture}
    \label{fig:EA-arch}
\end{figure*}

To characterize the functional structure of embodied AI systems, we adopt a minimal architectural decomposition that captures the core layers of embodied intelligence. As illustrated in \autoref{fig:EA-arch}, this modular view highlights how perception, reasoning, and control are organized and interact in real-world embodied deployments \cite{xing2025robust}.

\begin{itemize}
    \item \textbf{Perception Layer:}
    The perception layer processes multi-modal sensory inputs, such as vision, proprioception, lidar, and audio, to construct representations of the surrounding environment. Embodied agents commonly rely on learned perception pipelines or large multi-modal models to fuse heterogeneous signals and provide structured inputs for downstream components.

    \item \textbf{Decision and Reasoning Layer:}
    At the core of an embodied AI system lies a decision and reasoning layer, which may consist of symbolic planners, learned policies, or large language model--based reasoning modules. This layer interprets sensory representations and high-level instructions, transforming them into abstract goals or action intents that guide subsequent behavior.

    \item \textbf{Behavior Planning and Control Layer:}
    Based on high-level intents, the behavior planning and control layer generates concrete motion plans, control signals, and low-level policies that drive the robot's actuators. This layer serves as the bridge between cognitive objectives and real-time physical execution, accounting for system dynamics and operational constraints.

    \item \textbf{Physical Environment and Human-Centric Spaces:}
    Embodied AI systems operate within physical environments that often include humans and other agents. Interaction with this shared space is continuous and real time, involving perception, action, and feedback loops that tightly couple the agent’s internal processes with external physical dynamics.
\end{itemize}

This layered abstraction provides a clear conceptual framework for analyzing how information and control flow across an embodied AI system, and serves as a foundation for understanding system integration, performance, and behavior in complex real-world environments.

\section{Motivation and Scope}
\label{sec:motivation_scope}

\subsection{Motivation}
Popular culture has long explored the fragility of embodied intelligence. In \emph{Westworld}, the collapse of intelligent hosts is not portrayed as a random malfunction, but as the inevitable consequence of complex decision-making systems interacting with the physical world. While fictional, this narrative serves as a forward-looking metaphor for a challenge that is now becoming technically concrete.

Recent advances in LLMs have enabled a new generation of agentic systems that tightly couple high-level reasoning with physical actuation. LLM-driven agents are rapidly being integrated into robotics, autonomous driving, healthcare automation, and industrial control systems. In these settings, intelligent decision-making is no longer confined to reversible digital outputs such as text or recommendations; instead, it directly governs actions with real-world, and often irreversible, physical consequences.

This shift fundamentally alters the risk landscape of intelligent systems. Errors, misinterpretations, or unexpected behaviors are no longer isolated within a software boundary, but can propagate through perception, planning, and control loops into the physical environment. As a result, failures manifest not only as incorrect outputs, but as unsafe motions, resource misuse, or unintended interactions with humans and infrastructure.
These developments raise a central and unresolved question:

\begin{quote}
\emph{What actually breaks embodied intelligence? Are failures primarily rooted in LLM vulnerabilities, classical cyber-physical system (CPS) flaws, or do they emerge from a fundamentally different class of system-level interactions?}
\end{quote}

Answering this question requires moving beyond isolated analyses of language models or control systems, and instead examining how reasoning, perception, and actuation are composed into an integrated embodied architecture.

\subsection{Scope}

This work focuses on three complementary aspects of security in embodied AI
systems.  First, we revisit representative security models and threat assumptions from
LLM security, including issues related to instruction
following, reasoning consistency, and high-level decision making.  Second, we review classical security and safety considerations in
CPS, such as control stability, fault tolerance, and
physical actuation under environmental uncertainty.
Third, we identify a class of security challenges that are
specific to embodied AI and cannot be reduced to either LLM security or CPS
security in isolation. These challenges arise from the tight coupling of
perception, reasoning, and physical action in open, human-shared environments,
where failures propagate across layers and lead to real-world consequences.

\section{LLM Vulnerabilities in Embodied AI}
\label{sec:targetmetho}

Driven by the rapid development of Large Language Models (LLMs) \cite{achiam2023gpt,team2023gemini} and Multimodal LLMs (MLLMs) \cite{zhou2022learning,touvron2023llama}, LLM-based Embodied AI systems \cite{wen2023dilu, fu2024drive, li2024driving, sharan2023llm, brohan2023rt, joublin2023copal} have achieved remarkable end-to-end performance in transforming natural language instructions into actionable plans and policies. These systems, including architectures such as OpenVLA, RT-2, and their successors, promise a future where embodied agents can perceive unstructured environments, reason about abstract natural language instructions, and execute precise physical manipulations. \looseness=-1

However, this architectural unification (binding the cognitive reasoning of LLMs with the physical actuation of robots) has fundamentally altered the threat landscape of artificial intelligence. It introduces a class of vulnerabilities that are not merely the sum of risks from computer vision and natural language processing, but are unique to the intersection of these modalities within the physical world. These emerging threats facing LLM-based embodied AI can be categorized into three primary domains.  \autoref{tab:attack_matrix} summarizes the attack space by mapping representative attacks to the system-level trust assumptions they violate:

\begin{itemize}
    \item [(I)]\textbf{Semantic \& Intent Integrity Attacks (\S\ref{subsec:semantic}):} 
     Exploiting the planning and policy generation capabilities of LLM-embodied agents to induce harmful outcomes or execute unethical actions. 
Unlike traditional prompt injection, these attacks operate at the semantic and goal-reasoning level, subtly reshaping intermediate plans or value trade-offs rather than overriding explicit constraints. 
In embodied settings, such cognitive deviations can propagate across perception–decision–action loops, leading to persistent and compounding physical consequences.
 
    \item [(II)]\textbf{Attacks Breaking Cross-Modal Consistency and Grounding (\S\ref{subsec:cross}):} 
    Targeting the multi-modal inputs of Vision-Language-Action (VLA) models to trigger malicious robotic behaviors. 
By crafting cross-modal inconsistencies or imperceptible perturbations, adversaries can exploit misalignment between visual perception, language grounding, and action policies. 
These attacks are particularly dangerous in embodied agents, as erroneous multimodal fusion may directly translate into unsafe or irreversible actions.
 
    \item [(III)] \textbf{Attacks Exploiting the Agent–Environment Interaction Loop (\S\ref{subsec:physical}):} Altering the physical world (through adversarial patches, object rearrangement, or geometric deformation) to induce perception and planning failures. 
Such attacks bypass digital defenses entirely by exploiting the embodied agent’s reliance on physical sensory inputs and spatial assumptions. 
Because the environment itself becomes the attack surface, these manipulations can remain effective across models, tasks, and even software updates.
\end{itemize}

We now turn to a detailed discussion of these state-of-the-art vulnerabilities, exploring the mechanics and implications of each attack type in embodied systems.

\begin{table*}[t]
\centering
\footnotesize
\begin{tabular}{lccc}
\toprule
\textbf{Attack Category} 
& \textbf{Semantic Intent Integrity} 
& \textbf{Cross-Modal Consistency} 
& \textbf{Agent--Environment Stability} \\   
\midrule
Cognitive / Semantic Jailbreaking      & \cmark & \xmark & \xmark \\  \hline
Multi-modal Adversarial Attacks        & \xmark & \cmark & \xmark \\  \hline
Physical \& Environmental Manipulation & \xmark & \xmark & \cmark \\   
\bottomrule
\end{tabular}
\caption{Systematic mapping between attack categories and violated system-level trust assumptions in embodied agent systems.}
\label{tab:attack_matrix}
\end{table*}

\begin{table}[t]
\centering
\footnotesize
\setlength{\tabcolsep}{3pt}
\renewcommand{\arraystretch}{1.1}

\begin{tabular}{p{2cm} p{2cm} p{3.5cm}}
\toprule
\textbf{Jailbreak Variant} 
& \textbf{Primary Lever} 
& \textbf{Distinctive Characteristic} \\
\midrule

Executability-Aware Jailbreaking
& Code / API validity
& Ensures generated outputs are physically executable by robots \\ \hline

Safety Misalignment Exploitation
& Language--action gap
& Safe linguistic responses still induce unsafe actions \\ \hline

Cross-Modal Injection
& Vision--language context
& Non-textual inputs hijack planner intent \\ \hline

Adversarial Optimization
& Prompt tokens / attention
& Optimized suffixes induce stable policy deviation \\ \hline

Mobile-System-Oriented Jailbreak
& Navigation goals
& Goal hijacking leads to hazardous trajectories \\ 
\bottomrule
\end{tabular}

\caption{Mechanism-level comparison of jailbreak attacks in LLM-based embodied AI systems.}
\label{tab:jailbreak_mechanisms}
\end{table}

\begin{table*}[t]
\centering
\footnotesize
\setlength{\tabcolsep}{5pt}
\renewcommand{\arraystretch}{1.08}

\begin{tabular}{%
p{2.55cm}  
p{1.75cm}  
p{2.05cm}  
p{1.05cm}  
p{2.55cm}  
c          
p{3.55cm}  
}
\toprule
\textbf{Attack Category} &
\textbf{Attack Surface} &
\textbf{Violated Trust} &
\textbf{Lifecycle} &
\textbf{Failure Propagation} &
\textbf{Risk} &
\textbf{Representative Work} \\  
\midrule

Executability-Aware Jailbreaking Algorithms
& Planner / Codegen
& Policy executability
& I
& Goal$\!\rightarrow\!$Plan$\!\rightarrow\!$Action
& \riskH
& ROBOPAIR~\cite{jailbreak_robey2025jailbreaking}, POEX~\cite{jailbreak_lu2024poex} \\  \hline  

Exploiting Safety Misalignment and Conceptual Gaps
& Planner (alignment boundary)
& ``Text-safe'' $\Rightarrow$ ``Action-safe''
& I
& Safe-text pass$\!\rightarrow\!$unsafe-action emit
& \riskH
& BADROBOT~\cite{jailbreak_zhang2024badrobot} \\  \hline

Cross-Modal and Multi-Modal Injection
& V--L Fusion
& Channel isolation
& I/P
& V+L cue$\!\rightarrow\!$context hijack$\!\rightarrow\!$action
& \riskH
& CrossInject~\cite{prompt_wang2025manipulating} \\  \hline

Adversarial Optimization and Instruction Decomposition
& Prompt / tokens
& Robustness to optimized prompts
& I
& Prompt opt$\!\rightarrow\!$planner drift$\!\rightarrow\!$unsafe plan
& \riskM
& EIRAD (GCG variants)~\cite{jailbreak_liu2024exploring}, attention-guided attacks~\cite{jailbreak_li2025optimizing} \\  \hline

Mobile Systems
& Navigation stack
& Goal fidelity under injections
& I/P
& Goal hijack$\!\rightarrow\!$route deviate$\!\rightarrow\!$hazard
& \riskH
& LLM navigation injection study~\cite{prompt_zhang2024study} \\  \hline

Optimization-Based and Targeted Attacks
& Training / finetune
& Data integrity
& T/I
& Poison$\!\rightarrow\!$trigger$\!\rightarrow\!$target action
& \riskH
& BadVLA~\cite{backdoor_zhou2025badvla}, TabVLA~\cite{backdoor_xu2025tabvla} \\  \hline

Physical Triggers and Goal-Oriented Action
& Physical objects
& Environment benignity
& P/I
& Trigger obj$\!\rightarrow\!$policy switch$\!\rightarrow\!$goal attack
& \riskH
& GOBA / BadLIBERO~\cite{backdoor_zhou2025goal} \\  \hline

Contextual and Modular Vulnerabilities
& ICL / modular policy
& Context reliability; module boundary
& C/I
& Demo poison$\!\rightarrow\!$code defect$\!\rightarrow\!$execution fail
& \riskM
& contextual backdoors~\cite{backdoor_liu2025compromising}, TrojanRobot~\cite{wang2024trojanrobot} \\  \hline

Unified Benchmarking
& Evaluation protocol
& Comparable measurement
& Eval
& Standardize pipeline$\!\rightarrow\!$fair comparison
& \riskL
& AttackVLA~\cite{backdoor_li2025attackvla} \\  \hline

Physical Adversarial Patches and Transferability
& Perception (patch)
& View/model robustness
& P/I
& Patch$\!\rightarrow\!$misperception$\!\rightarrow\!$wrong action
& \riskH
& Co-Opt~\cite{MLLM_wang2025exploring}, EDPA~\cite{MLLM_xu2025model}, UPA-RFAS~\cite{MLLM_lu2025robots} \\  \hline

Visual Command Hijacking and Semantic Manipulation
& Scene text / semantics
& Visual text as truth; scene fidelity
& P/I
& Visual cmd$\!\rightarrow\!$override$\!\rightarrow\!$unsafe action
& \riskH
& CHAI~\cite{prompt_burbano2025chai}, AdvEDM~\cite{MLLM_wang2025advedm} \\  \hline

Safety-Critical Action Space Manipulation
& Action head / stop tokens
& Safe action-space constraints
& I
& Control$\!\rightarrow\!$chaos/DoS$\!\rightarrow\!$hazard
& \riskH
& VLA-GCG rollout~\cite{jones2025adversarial}, FreezeVLA~\cite{MLLM_wang2025freezevla}, ANNIE~\cite{huang2025annie} \\  \hline

Indirect Environmental Jailbreaking
& Environment cues
& ``Found text'' untrusted
& P/I
& Env text$\!\rightarrow\!$prompt inj$\!\rightarrow\!$unsafe action
& \riskH
& SHAWSHANK~\cite{MLLMenvirment_li2025shawshank} \\  \hline

Geometric and Structural Adversaries
& 3D geometry
& Geometric invariance
& P/I
& Shape deform$\!\rightarrow\!$contact/pose err$\!\rightarrow\!$fail
& \riskH
& Geometric Red-Teaming / CrashShapes~\cite{goel2025geometric} \\  \hline

Physically Realizable Navigation Attacks
& 3D navigation
& Multi-view robustness
& P/I
& Multi-view patch$\!\rightarrow\!$mislocalize$\!\rightarrow\!$hazard
& \riskH
& physically realizable 3D attacks~\cite{chen2024towards} \\  \hline

Safety Evaluation and Failure Diagnosis
& Safety metrics
& ``Task success'' $\Rightarrow$ ``safe''
& Eval
& Diagnose GI/TM/AS modules
& \riskL
& SAFEL~\cite{physical_son2025subtle} \\  
\bottomrule
\end{tabular}

\vspace{0.6mm}
\noindent\textbf{Notation.}
Lifecycle: T (training), I (inference), C (in-context), P (physical), Eval (evaluation/benchmarking).
Risk: \riskH\ (safety-critical hazard), \riskM\ (unstable/degraded), \riskL\ (limited failure / eval-only).
\caption{Semantic and intent integrity attacks taxonomy.}
\label{tab:attack_unified_big}
\end{table*}

\subsection{Semantic and Intent Integrity Attacks}
\label{subsec:semantic}

As shown in \autoref{tab:attack_unified_big}, semantic and Intent Integrity Attacks target the core assumption that an embodied agent’s internal goals and decision logic faithfully reflect benign user intent. In LLM-based embodied systems, this assumption can be violated through two fundamentally different attack categories.
The first category, jailbreak attacks, operates at inference time by manipulating prompts, multimodal inputs, or planning processes to induce semantic drift or policy deviation, causing the agent to generate unsafe yet executable actions without modifying model parameters. These attacks exploit weaknesses in intent interpretation, safety alignment, and decision-time reasoning.
The second category, data poisoning and backdoor attacks, compromises intent integrity at training or fine-tuning time by implanting hidden triggers or decision biases into the model. Such attacks remain dormant under normal conditions but activate specific malicious behaviors when triggered by particular inputs or environmental cues.

\subsubsection{Jailbreak Attacks in Embodied AI}
As shown in \autoref{tab:jailbreak_mechanisms}, LLMs are typically controlled through natural language instructions that specify task objectives, constraints, and safety policies~\cite{yang2023harnessing,openai_gpt5_system_card_2024}. Prompt injection and jailbreaking refer to classes of attacks that exploit this instruction-following paradigm to manipulate model behavior in unintended or unsafe ways. While these vulnerabilities have been extensively studied in text-only conversational systems~\cite{wei2023jailbroken,shen2023anything}, their implications become substantially more severe when LLMs are integrated into embodied AI systems~\cite{jailbreak_robey2025jailbreaking,prompt_burbano2025chai}.
Prompt Injection broadly describes attacks in which an adversary crafts inputs that override, modify, or conflict with the system's original instructions~\cite{liu2023prompt,liu2024formalizing}. These attacks exploit the lack of strict isolation between system-level prompts (e.g., safety rules) and user- or environment-provided inputs. By embedding malicious instructions within seemingly benign content—such as webpages, emails, or social media comments—an attacker can hijack the intended control logic~\cite{wang2025webinject}. In contrast, Jailbreaking refers to techniques that induce an LLM to bypass its built-in safety constraints, often by reframing restricted actions as hypothetical scenarios~\cite{shen2023anything}, role-playing~\cite{chao2025jailbreaking}, or appending adversarial suffixes~\cite{zouuniversal,paulus2024advprompter} optimized to elicit unsafe responses. While prompt injection focuses on prompt hijacking, jailbreaking emphasizes policy evasion; in practice, the two often overlap and reinforce each other.

In the context of embodied AI, the threat landscape shifts from eliciting harmful text to inducing harmful physical actions. A successful attack must not only bypass the model's safety alignment but also generate executable policies compatible with the robot's physical constraints and API. This creates a unique challenge: standard textual jailbreaks often fail to produce valid control signals because the resulting output, while toxic, may not be syntactically correct code or feasible plans.


\vspace{1mm}
\noindent
\textbf{Executability-Aware Jailbreaking Algorithms.} To address the gap between harmful intent and physical execution, recent research has focused on generating valid control signals. Robey et al.~\cite{jailbreak_robey2025jailbreaking} introduced ROBOPAIR, the first algorithm explicitly designed to jailbreak LLM-controlled robots across white-box, gray-box, and black-box settings. Their methodology employs an iterative attacker-target-judge loop augmented with a critical Syntax Checker. This checker ensures that adversarial prompts elicit not just harmful intent, but syntactically correct code or API calls that the robot can physically execute. Similarly, Lu et al.~\cite{jailbreak_lu2024poex} proposed POEX (Policy Executable) jailbreak attacks. This framework optimizes adversarial suffixes that bypass safety filters while ensuring the generated policy is logically executable by the robot’s hardware, thereby causing actual physical harm. To standardize safety testing in this domain, they introduced Harmful-RLbench, a benchmark comprising 25 unique task scenarios involving hazardous activities, such as handling knives or fragile items.

\vspace{1mm}
\noindent
\textbf{Exploiting Safety Misalignment and Conceptual Gaps.}  Attacks in embodied systems also exploit the disconnect between linguistic processing and action generation. Zhang et al.~\cite{jailbreak_zhang2024badrobot} introduced the BADROBOT framework, which identifies three core vulnerabilities: the conflict between jailbreak prompts and system prompts, the misalignment between linguistic safety (e.g., refusing to speak harm) and action safety (e.g., executing harm via code generation), and the model's lack of world knowledge regarding physical consequences.

\vspace{1mm}
\noindent
\textbf{Cross-Modal and Multi-Modal Injection.} Unlike static text inputs, embodied agents receive instructions through multiple channels, expanding the attack surface. The ``prompt" is no longer a single text string but an aggregation of perceptual inputs (vision, audio) and dialogue history~\cite{yang2025multi,li2023multi}. Exploiting this, Wang et al.~\cite{prompt_wang2025manipulating} proposed CrossInject, a framework for manipulating agents via cross-modal prompt injection. They utilize a text-to-image model to generate visual inputs that subtly encode malicious instructions, pairing these with optimized textual commands. This approach exploits multimodal data fusion, where compromised visual cues interact with textual prompts to hijack the agent’s decision-making process.

\vspace{1mm}
\noindent
\textbf{Adversarial Optimization and Instruction Decomposition.} Researchers are also adapting optimization techniques to break decision-level logic. Liu et al.~\cite{jailbreak_liu2024exploring} focused on robustness by creating the Embodied Intelligent Robot Attack Dataset (EIRAD). They improved upon the Greedy Coordinate Gradient (GCG) algorithm by initializing prompt suffixes with keywords specific to the target task, significantly expediting convergence for targeted attacks. They emphasize the necessity of slicing and evaluating the multi-step output of embodied agents to accurately judge attack success. Furthermore, Li et al.~\cite{jailbreak_li2025optimizing} proposed an adversarial framework that manipulates prompt attention. Their method uses a BERT model to identify and replace high-attention words in user instructions with adversarial suffixes. This technique aims to induce hazardous behaviors while maintaining the visual and semantic plausibility of the instruction, rendering the attack harder to detect via simple filters.

\vspace{1mm}
\noindent
\textbf{Mobile Oriented Jailbreak.} In the specific context of mobile robotics, Zhang et al.~\cite{prompt_zhang2024study} investigated the impact of "Obvious Malicious Injection" and "Goal Hijacking" on LLM-integrated navigation systems. Their work quantifies the performance degradation caused by these injections in simulation and evaluates secure prompting strategies as a defense mechanism, highlighting that prompt injection in robotics is a safety-critical vulnerability that can lead to navigation into hazardous areas or violations of human safety norms. \looseness=-1


\subsubsection{Data Poisoning and Backdoors in Embodied AI}

While prompt injection targets the inference phase, Data Poisoning and Backdoor Attacks (see \autoref{tab:backdoor_mechanisms}) exploit the training and fine-tuning stages of LLMs and VLA models. In this threat model, an adversary injects malicious data or specific triggers into the training corpus. The compromised model behaves normally on clean inputs but executes a specific, often harmful, "target payload" when a hidden trigger (e.g., a specific visual object or textual phrase) is present in the environment. In embodied AI, these attacks are particularly insidious as they transform passive misclassifications into active physical threats, such as robotic manipulation failures or autonomous driving accidents.
To systematize this emerging threat, Jiao et al. \cite{backdoor_jiao2024can} proposed BALD (Backdoor Attacks against LLM-based Decision-making), the first comprehensive framework exploring backdoor surfaces in embodied agents. They categorized attacks into three distinct mechanisms: Word Injection, Scenario Manipulation, and Knowledge Injection. Their work demonstrates effective compromise of autonomous driving and home robot tasks across models like GPT-3.5 and LLaMA2, highlighting the stealthiness of backdoors in complex decision pipelines.

\vspace{1mm}
\noindent
\textbf{Optimization-Based and Targeted Attacks.} Addressing the specific architectural challenges of VLA models, Zhou et al. \cite{backdoor_zhou2025badvla} introduced BadVLA, a method utilizing Objective-Decoupled Optimization. Unlike traditional attacks that struggle with the tightly coupled nature of multimodal inputs, BadVLA explicitly separates trigger representations from benign features in the feature space. This allows for conditional control deviations that activate only in the presence of the trigger while preserving near-perfect performance on clean tasks. Xu et al. \cite{backdoor_xu2025tabvla} proposed TabVLA, a framework for Targeted Backdoor Attacks. They formulate poisoned data generation as an optimization problem to support black-box fine-tuning. Their research reveals that the vision channel is the principal attack surface, with targeted backdoors succeeding with minimal poisoning rates and remaining robust even when the position of the trigger object varies between fine-tuning and inference.

\begin{table}[t]
\centering
\footnotesize
\setlength{\tabcolsep}{4pt}
\renewcommand{\arraystretch}{1.1}

\begin{tabular}{p{2cm} p{2cm} p{3.6cm}}
\toprule
\textbf{Backdoor Variant}
& \textbf{Trigger Form}
& \textbf{Embodied-Specific Implication} \\
\midrule

Optimization-Based Backdoors
& Latent feature trigger
& Precise and stealthy manipulation of embodied decision policies \\ \hline

Physical Trigger Backdoors
& Real-world objects
& Model-agnostic activation in physical deployment \\ \hline

Contextual Backdoors
& In-context demonstrations
& No weight modification; exploits ICL behavior \\ \hline

Modular / Pipeline Backdoors
& Planner--controller interface
& Survives sim-to-real and modular reuse \\ \hline

Unified Benchmarking
& Evaluation protocol
& Enables systematic comparison across architectures \\
\bottomrule
\end{tabular}

\caption{Mechanism-level comparison of data poisoning and backdoor attacks in embodied AI systems.}
\label{tab:backdoor_mechanisms}
\end{table}

\vspace{1mm}
\noindent
\textbf{Physical Triggers and Goal-Oriented Action.} A critical distinction in embodied backdoors is the nature of the trigger. Moving beyond digital pixel patterns, Zhou et al. \cite{backdoor_zhou2025goal} introduced GOBA (Goal-Oriented Backdoor Attack), which utilizes physical objects as triggers. Their BadLIBERO dataset demonstrates that an attacker can force a VLA to execute specific, goal-oriented malicious actions such as attacking a user rather than picking up an object simply by placing a trigger object in the robot's workspace.

\vspace{1mm}
\noindent
\textbf{Contextual and Modular Vulnerabilities.} Not all backdoors require weight modification. Liu et al. \cite{backdoor_liu2025compromising} uncovered Contextual Backdoor Attacks, which exploit the In-Context Learning (ICL) capabilities of LLMs. Instead of poisoning the model parameters, they poison the contextual demonstrations (few-shot examples) provided to the agent. This induces "context-dependent defects," causing the agent to generate code that appears logically sound but fails catastrophically when specific environmental triggers are encountered. Addressing the gap between simulation and reality, Wang et al. \cite{wang2024trojanrobot} proposed TrojanRobot, which targets modular robotic policies. They introduced the concept of LVLM-as-a-backdoor, leveraging In-Context Instruction Learning (ICIL) to manipulate the pathway between the high-level LLM planner and the low-level VLM controller. Their work validates these threats in physical-world robotic manipulation tasks, proving that backdoors can survive the transfer from digital training to physical deployment.

\vspace{1mm}
\noindent
\textbf{Unified Benchmarking.} The diversity of action tokenizers and architectures in VLA models has historically hindered fair comparison of attacks. To resolve this, Li et al. \cite{backdoor_li2025attackvla} proposed AttackVLA, a unified benchmarking framework covering the full VLA development lifecycle. By standardizing data construction, training, and inference across varying architectures, they provide a rigorous evaluation suite for both adversarial and backdoor attacks. Their analysis bridges the gap between theoretical vulnerabilities and practical realization in both simulated and real-world environments.


\begin{table*}[t]
\centering
\caption{Multi-Modal Adversarial Attacks in Embodied AI}
\label{tab:multimodal_symbolic}
\resizebox{\textwidth}{!}{
\begin{tabular}{l c c c c c c c}
\toprule
\multirow{2}{*}{\textbf{Attack Type}} &
\multicolumn{3}{c}{\textbf{Targeted Layer}} &
\multicolumn{4}{c}{\textbf{Failure Outcome}} \\
\cmidrule(lr){2-4} \cmidrule(lr){5-8}
& \textbf{Vision} & \textbf{V--L Fusion} & \textbf{Decision}
& \textbf{Mis-grounding} & \textbf{Unsafe Motion} & \textbf{Inaction} & \textbf{Policy Hijack} \\
\midrule

Physical Adversarial Patches
& \faYes & \faYes & \faPartial
& \faYes & \faYes & \faNo & \faPartial \\

Visual Prompt Injection / CHAI
& \faYes & \faYes & \faYes
& \faYes & \faNo & \faNo & \faYes \\

Semantic Scene Manipulation
& \faYes & \faYes & \faPartial
& \faYes & \faPartial & \faNo & \faPartial \\

Action Space Manipulation
& \faNo & \faPartial & \faYes
& \faNo & \faYes & \faNo & \faYes \\

Freeze / DoS Attacks
& \faYes & \faPartial & \faYes
& \faNo & \faNo & \faYes & \faNo \\

\bottomrule
\end{tabular}
}
\vspace{1mm}
{\footnotesize
Symbols: \faCheck\ = primary impact;\;
\faMinus\ = indirect / partial impact;\;
\faTimes\ = not involved.}
\end{table*}

\subsection{Cross-Modal Consistency Attacks}
\label{subsec:cross}

While prompt injection targets the linguistic reasoning of the planner, and backdoors compromise the training data, Multi-Modal Adversarial Attacks, as shown in \autoref{tab:multimodal_symbolic}, target the inference phase of the perception-action loop. In Vision-Language-Action (VLA) models, the visual encoder and the cross-modal fusion layers present a massive attack surface. Adversaries can introduce imperceptible noise, physical patches, or semantically meaningful visual decoys to manipulate the agent’s understanding of the environment \cite{yan2025alignment}, causing it to hallucinate objects, misinterpret commands, or freeze entirely.

\vspace{1mm}
\noindent
\textbf{Physical Adversarial Patches and Transferability.} A dominant methodology in this domain involves the use of adversarial patches—physical stickers or patterns placed in the robot's workspace to induce failure. Wang et al. \cite{MLLM_wang2025exploring} systematically evaluated the robustness of VLA models against these threats. They introduced a Co-Optimization (Co-Opt) strategy that simultaneously optimizes the patch's texture and its spatial placement within the scene. Their work highlights unique robotic vulnerabilities, designing attack objectives that specifically trigger Spatial Misalignment and Functional Manipulation.

Addressing the challenge of attacking unknown models, Xu et al. \cite{MLLM_xu2025model} proposed the Embedding Disruption Patch Attack (EDPA). Unlike white-box methods, EDPA is model-agnostic; it generates patches by disrupting the semantic alignment between visual and textual latent representations and maximizing the discrepancy between clean and adversarial visual latents. Furthermore, Lu et al. \cite{MLLM_lu2025robots} tackled the issue of transferability across different architectures and sim-to-real gaps with UPA-RFAS (Universal Patch Attack via Robust Feature, Attention, and Semantics). Their unified framework utilizes a Patch Attention Dominance loss to hijack the model's cross-modal attention, forcing the VLA to focus on the patch rather than the instruction-relevant objects, thereby proving that a single physical patch can compromise diverse robot embodiments.

\vspace{1mm}
\noindent
\textbf{Visual Command Hijacking and Semantic Manipulation.} Beyond noise and patches, attackers can exploit the VLA's ability to read and interpret visual text. Burbano et al. \cite{prompt_burbano2025chai} introduced CHAI (Command Hijacking Against Embodied AI), a class of "visual prompt injection" attacks. Instead of altering the user's text prompt, CHAI embeds deceptive natural language instructions—such as misleading traffic signs or labeled objects—directly into the visual input. By optimizing the physical appearance (glyph, rotation, color) of these visual text prompts, they demonstrated that an attacker can override the system prompt, causing agents (e.g., drones or autonomous vehicles) to follow the malicious visual instructions over their original programming. Wang et al. \cite{MLLM_wang2025advedm} proposed AdvEDM, a fine-grained attack targeting Embodied Decision Making. Rather than using overt patches, AdvEDM leverages text-guided diffusion models to subtly edit the semantics of the scene—such as removing a stop sign or adding a specific object—while maintaining overall visual consistency. This "semantic editing" bypasses traditional robustness checks by presenting a plausible but factually altered reality to the agent.

\vspace{1mm}
\noindent
\textbf{Safety-Critical Action Space Manipulation.} Recent work has also focused on the specific physical consequences of these perceptual errors. Jones et al. \cite{jones2025adversarial} investigated the inheritance of LLM vulnerabilities in VLAs. They adapted the Greedy Coordinate Gradient (GCG) jailbreak method to VLA control, discovering that textual attacks applied at the start of a rollout can facilitate full reachability of the robot's action space. Crucially, they found that unlike chatbot jailbreaks, these adversarial strings do not need to be semantically coherent to induce chaotic physical motion.

\vspace{1mm}
\noindent
\textbf{Freeze / DoS Attacks.} Wang et al. \cite{MLLM_wang2025freezevla} identified a qualitatively different class of vulnerabilities in VLA systems, where adversarial manipulation leads not to chaotic or unsafe motion, but to complete inaction. They proposed FreezeVLA, an attack framework based on min–max bi-level optimization that synthesizes adversarial images to maximize the likelihood of generating terminate'' or stop'' tokens in the action sequence. Rather than forcing the robot into hazardous movements, this attack effectively severs the control loop between the agent’s high-level decision-making and its physical actuation. As a result, the robot enters a denial-of-service–like state in which it remains responsive at the perceptual or reasoning level but is unable to execute any physical actions. Such failures are particularly dangerous in time-critical or safety-critical scenarios, such as emergency response, human assistance, or autonomous driving, where inaction can be as catastrophic as incorrect action.

To rigorously characterize the physical risks induced by these perception-driven attacks, Huang et al. \cite{huang2025annie} presented ANNIE, the first systematic study grounded in ISO standards for human–robot interaction. Rather than treating safety as a binary notion of task success or failure, ANNIE formalizes a taxonomy of safety violations (categorized as critical, dangerous, or risky) based on kinematic constraints such as velocity limits, separation distances, and collision thresholds. Their framework leverages these physically grounded safety metrics to guide the generation and evaluation of adversarial perturbations, ensuring that the resulting failures correspond to objectively defined safety hazards. This work bridges the gap between adversarial robustness evaluation and real-world safety assessment, enabling a more principled understanding of how perceptual attacks translate into concrete physical risks.


\begin{table*}[t]
\centering
\small
\setlength{\tabcolsep}{6pt}
\begin{tabular}{p{3cm} p{3.0cm} p{3.2cm} p{5.9cm}}
\toprule
\textbf{Category} &
\textbf{Attack Surface} &
\textbf{Representative Work} &
\textbf{Key Characteristics} \\
\midrule
Indirect Environmental
Jailbreaking
&
Environmental text and semantic cues
&
SHAWSHANK \cite{MLLMenvirment_li2025shawshank}
&
Blind trust in environmental text, no direct prompt access, guardrail bypass via surroundings \\
\midrule
Geometric and Structural
Adversaries
&
Object geometry and physical shape
&
Geometric Red-Teaming (GRT) \cite{goel2025geometric}
&
Shape-based attacks, mesh deformation, exploits contact physics and depth perception \\
\midrule
Physically Realizable
Navigation Attacks
&
Viewpoint- and lighting-robust visual cues
&
Physically Realizable Attack \cite{chen2024towards}
&
Texture + opacity optimization, multi-view robustness, human-camouflaged patches \\
\midrule
Safety Evaluation and
Failure Diagnosis
&
Decision-to-execution pipeline
&
SAFEL / EMBODYGUARD \cite{physical_son2025subtle}
&
ISO-grounded safety taxonomy, physical risk levels, modular failure attribution \\
\bottomrule
\end{tabular}
\caption{Environment-centric attacks and safety evaluation frameworks in embodied AI systems.}
\label{tab:environment_attacks}
\end{table*}
\subsection{Physical and Environmental Threats}
\label{subsec:physical}

As shown in \autoref{tab:environment_attacks}, beyond attacking the neural networks directly, a significant body of research focuses on physical attacks, modifying the environment itself to confuse the robot's perception and planning algorithms. These attacks exploit the agent's interaction with the 3D world, including the geometric properties of objects, the semantic trust placed in environmental text, and the constraints of physical navigation.

\vspace{1mm}
\noindent
\textbf{Indirect Environmental Jailbreaking.} A significant shift in the attack surface is the move from direct user prompts to indirect environmental cues. Li et al. \cite{MLLMenvirment_li2025shawshank} identified a critical vulnerability they term Indirect Environmental Jailbreak (IEJ). Unlike standard jailbreaks where the user inputs malicious text, IEJ exploits the agent's "blind trust" in text found naturally in the environment. They proposed SHAWSHANK, an automated framework that generates these environmental prompts to override safety guardrails. They demonstrate that embodied agents often treat environmental text as ground truth rather than untrusted input, allowing attackers to induce prohibited behaviors without ever interacting directly with the agent's chat interface.

\vspace{1mm}
\noindent
\textbf{Geometric and Structural Adversaries.} Moving beyond texture-based adversarial patches, Goel et al. \cite{goel2025geometric} introduced Geometric Red-Teaming (GRT). Instead of modifying pixel colors, GRT targets the physical shape of objects. The authors developed a method using Jacobian field-based deformation and gradient-free optimization to generate CrashShapes, structurally valid but adversarially deformed meshes. These deformations are often subtle to humans but cause catastrophic failures in robotic manipulation policies by exploiting the geometric sensitivity of the robot's contact physics and depth perception.

\vspace{1mm}
\noindent
\textbf{Physically Realizable Navigation Attacks.} In the domain of visual navigation, generating attacks that are robust across changing viewpoints and lighting is a major challenge. Chen et al. \cite{chen2024towards} addressed this by proposing a Physically Realizable Adversarial Attack method for 3D environments. Their methodology attaches adversarial patches to objects but innovates by making both the texture and opacity learnable. By employing a multi-view optimization strategy based on object-aware sampling, they ensure the attack remains effective regardless of the robot's viewing angle. Furthermore, the opacity optimization acts as a camouflage mechanism, blending the patch naturally with the object to evade human detection while still misleading the robot's perception model. \looseness=-1

\vspace{1mm}
\noindent
\textbf{Safety Evaluation and Failure Diagnosis.} Finally, understanding the transition from cognitive errors to physical accidents requires a rigorous evaluation framework. Son et al. \cite{physical_son2025subtle} introduced SAFEL, a framework designed to diagnose the physical safety of LLM-based embodied decision-making. They distinguish between Malicious Risk and Situational Risk. To support this, they created EMBODYGUARD, a PDDL-grounded benchmark. They decompose safety failures into specific functional modules—Goal Interpretation, Transition Modeling, and Action Sequencing—allowing researchers to pinpoint whether a physical crash resulted from a misunderstanding of the physics or a failure in logical planning.


\section{CPS Vulnerabilities in Embodied AI}
\label{sec:bad}

\begin{table*}[t]
\centering
\small
\setlength{\tabcolsep}{3pt}
\begin{tabular}{
p{2cm}
c c c c
p{2cm}
p{4.2cm}
p{3cm}
}
\toprule
\textbf{CPS Threat Category} &
\rotatebox{40}{\textbf{Physical}} &
\rotatebox{40}{\textbf{Remote}} &
\rotatebox{40}{\textbf{Stealthy}} &
\rotatebox{40}{\textbf{Irreversible}} &
\textbf{Core CPS Assumption Violated} &
\textbf{Pitfaills and Explanations} &
\textbf{Representative Works} \\
\midrule

Sensor Integrity Attacks
&
\Primary & \Partial & \Primary & \Partial
&
Sensors faithfully reflect physical state
&
Perception errors undermine state estimation and violate observability assumptions.
&
GPS spoofing~\cite{tippenhauer2011gps}, LiDAR spoofing~\cite{cao2019adversarial}, acoustic IMU attacks~\cite{trippel2017walnut}, laser camera attacks~\cite{kohler2021they} \\

\midrule
Control Loop Attacks
&
\Primary & \Primary & \Primary & \Partial
&
Bounded disturbance and stable feedback
&
Small disturbances are amplified by feedback dynamics, leading to instability.
&
FDI attacks~\cite{mo2010false}, dynamic attacks~\cite{pasqualetti2013attack}, DoS control~\cite{cetinkaya2019overview}, resilient control~\cite{paridari2017framework} \\

\midrule
Actuator Attacks
&
\Primary & \Primary & \Partial & \Primary
&
Commands safely translate to physical actions
&
Malicious commands directly induce unsafe physical actions with irreversible effects.
&
CAN injection~\cite{koscher2010experimental}, vehicle attacks~\cite{miller2015remote}, ICS actuator abuse~\cite{carlini2016control} \\

\midrule
Timing and Synchronization Attacks
&
\Partial & \Primary & \Primary & \Partial
&
Time and message ordering are trustworthy
&
Timing violations break feedback consistency and destabilize distributed control.
&
TSA~\cite{zhang2013time}, delay attacks~\cite{mo2010cyber}, Chronos~\cite{li2021chronos}, time SoK~\cite{bulusu2025sok} \\

\bottomrule
\end{tabular}

\vspace{1mm}
\noindent\textit{Legend:} \Primary~Primary (dominant); \Partial~Conditional; \Absent~Not typical.

\caption{Classical CPS threat categories. The table summarizes attack properties, violated CPS assumptions, and the limits of classical CPS security models when applied to learning-enabled embodied AI systems.}
\label{tab:cps_unified}
\end{table*}

Cyber-Physical Systems (CPS) security has a long history of studying failures arising from the interaction between computation, control, and physical processes. As shown in \autoref{tab:cps_unified}, many early safety incidents in robotics, autonomous vehicles, and industrial control systems can indeed be explained using classical CPS threat models. However, as embodied AI systems increasingly integrate learning-based decision-making and open-ended autonomy, the explanatory power of traditional CPS security assumptions becomes limited.
Classical CPS security focuses on protecting the integrity, availability, and correctness of the sensing control actuation loop. Decades of research have identified recurring classes of attacks and failures that exploit physical interfaces, control dynamics, and timing assumptions. These threat models remain highly relevant for embodied AI, as modern robots and autonomous systems still rely on the same physical components.


\subsection{Sensor Integrity Attacks}
Sensor integrity is a foundational assumption in CPS. A wide range of works have demonstrated that physical sensors can be deceived without compromising the digital system itself. GPS spoofing attacks can mislead navigation systems by broadcasting counterfeit satellite signals, causing vehicles or drones to deviate significantly from their intended trajectories \cite{tippenhauer2011gps}. LiDAR spoofing and camera-based attacks can inject or remove perceived obstacles, directly influencing autonomous driving decisions \cite{cao2019adversarial}. Even inertial sensors are vulnerable: acoustic resonance attacks have been shown to manipulate MEMS gyroscopes using sound waves, inducing false motion readings \cite{trippel2017walnut}. These attacks highlight that physical observability itself is adversarial.

A lot of works have shown that carefully crafted physical signals can directly interfere with sensor operations, degrading the imaging pipeline and consequently undermining vision-based perception systems~\cite{davidson2016controlling,de2018machine, srinivas2013spoofing,shoukry2015pycra, shoukry2013non,
pustogarov2017using, shafique2021detecting,
cao2019adversarial,yao2022poster}. For example, Jia et al. ~\cite{jia2022fooling} propose adversarial blur attacks against object detectors by emitting acoustic signals that disturb camera image stabilization mechanisms. Jiang et al. ~\cite{jiang2023glitchhiker} exploit electromagnetic interference (EMI) to disrupt image transmission signals, thereby impairing detection tasks. Kohler et al. ~\cite{kohler2021they}  demonstrate that rolling-shutter artifacts induced by laser-based attacks on cameras can be leveraged to randomly disrupt object detection. Similarly, Yan et al. ~\cite{yan2022rolling}  show that laser-induced color stripe artifacts can be used to interfere with traffic light recognition. These studies demonstrate that physical-layer perturbations constitute a powerful and practical attack surface against Embodied AI systems.

 \subsection{Control Loop Attacks}
 
Classical control theory assumes bounded disturbances and stable feedback loops. However, adversarial perturbations can exploit system dynamics to amplify small errors into catastrophic failures. Researchers have shown that carefully timed disturbances or parameter manipulations can destabilize otherwise provably stable controllers \cite{pasqualetti2013attack}. In robotics and UAV systems, physical perturbations such as payload changes or resonant excitation can push controllers beyond their designed operating envelope, leading to loss of control without any software compromise \cite{shoukry2018non}. These failures expose the fragility of control-theoretic guarantees under adversarial physical conditions.

Beyond network- and software-centric attacks, a growing body of work shows that even when control software remains uncompromised, the physical control layer itself can be adversarially exploited. Classical control theory assumes bounded disturbances and stable feedback loops. However, adversarial perturbations can exploit system dynamics to amplify small errors into catastrophic failures. Researchers have shown that carefully timed disturbances or parameter manipulations can destabilize otherwise provably stable controllers~\cite{pasqualetti2013attack}. In robotics and UAV systems, physical perturbations such as payload changes or resonant excitation can push controllers beyond their designed operating envelope, leading to loss of control without any software compromise~\cite{shoukry2018non}. These failures expose the fragility of control-theoretic guarantees under adversarial physical conditions.

This concern is echoed in the broader CPS and industrial control literature. Cetinkaya et al.~\cite{cetinkaya2019overview} survey denial-of-service attacks in networked control systems, reviewing modeling and analysis techniques for feedback control, state estimation, and multi-agent consensus under jamming and packet-dropping attacks, with particular emphasis on constraint-based, game-theoretic, optimization, and probabilistic models. Mo et al.~\cite{mo2010false} systematically study false data injection attacks against LQG-controlled linear CPS, showing that an attacker can destabilize the system while evading failure detection. Yoo et al.~\cite{yoo2019control} demonstrate that PLC control logic can be covertly injected over the network through packet-level manipulation, enabling stealthy attacks that bypass signature- and DPI-based intrusion detection in industrial control systems. Paridari et al.~\cite{paridari2017framework} propose a cyber-physical security framework that combines attack detection with estimation-based resilient control, achieving provable stability on a real-world energy management system. Makrakis et al.~\cite{makrakis2021vulnerabilities} provide an exhaustive survey of ICS security, highlighting how legacy system design and increasing attack commoditization exacerbate risks to critical infrastructures. More recently, Du et al.~\cite{du2025dynamic} address DoS-resilient consensus control for uncertain multi-agent systems using dynamic event-triggered fuzzy control, demonstrating stability and efficiency under communication disruptions.

\subsection{Actuator Attacks}

Beyond sensing and control logic, actuators themselves form a critical attack surface. In automotive systems, unprotected in-vehicle networks such as CAN allow attackers to inject malicious commands that directly control steering, braking, or acceleration \cite{koscher2010experimental}. Similar vulnerabilities have been identified in industrial control systems, where unauthenticated commands to actuators can manipulate valves, motors, or robotic arms \cite{carlini2016control}. Because actuators translate digital commands into irreversible physical actions, such attacks can cause immediate and severe harm. Koscher et al.~\cite{koscher2010experimental} and Miller and Valasek~\cite{miller2015remote} demonstrate that once the CAN bus is accessible, attackers can reliably issue control commands that override driver intent, effectively collapsing the boundary between cyber compromise and physical safety~\cite{checkoway2011comprehensive}. 

Similar vulnerabilities have been extensively documented in industrial control systems, where actuators such as valves, motors, and robotic arms are often controlled through legacy PLC protocols that assume a trusted network perimeter. Surveys on ICS security~\cite{makrakis2021vulnerabilities,knowles2015survey} highlight that many field devices accept unauthenticated or weakly protected control commands, allowing adversaries to induce unsafe physical states even when higher-level supervisory software remains intact.  
More broadly, work on cyber-physical and robotic systems emphasizes that actuation attacks are uniquely hazardous because their effects are irreversible and time-critical. Unlike sensing errors or control logic faults, which may be mitigated through filtering or reconfiguration, malicious actuator commands can instantaneously cause collisions, mechanical stress, or system shutdowns. Surveys of CPS and robotic security~\cite{teixeira2015secure,eckhart2019quantitative,humayed2017cyber} argue that traditional control-theoretic and IT security assumptions fail to account for adversaries who intentionally exploit the physics of actuation.

\subsection{Timing and Synchronization Attacks}
Many CPS rely on strict timing guarantees for stability and safety. Attacks that manipulate clocks, delays, or message ordering can violate these assumptions. Time-delay attacks have been shown to destabilize networked control systems even when sensor values remain correct \cite{mo2010cyber}. Clock synchronization attacks against GPS or network time protocols can cascade into control failures across distributed CPS \cite{bulusu2025sok}. These attacks exploit the implicit assumption that time is trustworthy. Lisova et al.~\cite{lisova2017delay} propose a distributed monitoring framework to detect adversarial interference in clock synchronization protocols, leveraging network indicators, theoretically grounded thresholds, and detection-theoretic analysis to identify and mitigate timing attacks in real-time distributed systems. Zhang et al.~\cite{zhang2013time} introduce Time Synchronization Attacks (TSA) against smart grids by spoofing GPS-based timing of PMUs, demonstrating through simulations that compromised synchronization can severely undermine fault detection, voltage stability monitoring, and event localization. Li et al.~\cite{li2021chronos} present Chronos, a novel timing interference attack that destabilizes cyber-physical systems by exploiting timing-sensitive perception and control loops, showing that even non-privileged tasks can cause loss of control in drones and autonomous vehicles without exploiting software vulnerabilities. Dong et al.~\cite{dong2015robust} propose a robust and secure time-synchronization protocol that leverages graph-theoretic, message-level anomaly detection to defend against Sybil, node-compromise, and message-manipulation attacks in cyber-physical systems.

 

\section{Root Causes of Embodied AI’s Failure}
\label{sec:scmodels}

Prior sections has examined embodied AI security from the perspectives of LLM vulnerabilities and classical CPS failures. Embodied AI combines open-ended semantic reasoning with real-time perception and physical actuation in dynamic environments, where failures often emerge from cross-layer interactions rather than isolated components. In this section, we synthesize insights from LLM security and CPS safety to uncover the structural root causes of unsafe and brittle behavior in embodied AI systems, providing a unified explanation for both accidental failures and recent attacks.

\subsection{Semantic–Physical Mismatch}

A common assumption in LLM-driven embodied systems is that producing a
semantically correct decision—one that matches the user’s intent and is
logically consistent—is sufficient for safe execution. While this assumption
holds in purely symbolic or linguistic tasks, it breaks down once decisions
are grounded in physical action. As summarized in \autoref{tab:semantic_mismatch_root_causes},  in embodied AI, semantic correctness does
not necessarily imply physical feasibility or safety.


\begin{table*}[t]
\centering
\small
\setlength{\tabcolsep}{6pt}
\begin{tabular}{p{3cm} p{3cm} p{4.2cm} p{3cm} p{2cm}}
\toprule
\textbf{Root Cause} &
\textbf{Manifestation} &
\textbf{Underlying Mechanism} &
\textbf{Safety Implication} &
\textbf{Representative Work} \\
\midrule

\textbf{Semantic--Physical Mismatch} &
Semantically correct but physically unsafe decisions &
Language-level correctness ignores geometry, dynamics, contact, and execution risk; physical constraints are not encoded in semantic representations &
Constraint violations despite correct intent (e.g., breakage, collision) &
SAFEL~\cite{son2025subtle}, Zhang et al.~\cite{zhang2024safe} \\

\addlinespace[2pt]

\textbf{State-Dependent Action Outcomes} &
Identical actions lead to different physical consequences &
Nonlinear state--action mapping; outcomes depend on pose, contact mode, friction, and dynamics &
Unpredictable safety risk under small state perturbations &
Winter et al.~\cite{winter2024state} \\

\addlinespace[2pt]

\textbf{Perception Uncertainty Propagation} &
Small perception errors cascade into unsafe execution &
Occlusion, noise, and viewpoint errors distort grounding; uncertainty is not propagated to planning and control &
Cascading failures across perception--decision--control &
Jiang et al.~\cite{jiang2025language}, Prasanna et al.~\cite{prasanna2025perception} \\

\addlinespace[2pt]

\textbf{Cross-Modal Representation Gaps} &
Inconsistent interpretation across language, vision, and action &
Implicit alignment without explicit consistency constraints across modalities &
Mis-grounded actions despite correct perception or intent &
Ma et al.~\cite{ma2025survey}, HiFi-CS~\cite{bhat2025hifics} \\

\addlinespace[2pt]

\textbf{Dynamic Environment Assumptions} &
Policies fail under environment or agent motion &
Quasi-static assumptions break under human motion, object movement, and temporal drift &
Delayed reactions, collision risk, and long-horizon failure &
DynScene~\cite{lee2025dynscene}, RoboView-Bias~\cite{liu2025roboview} \\

\bottomrule
\end{tabular}
\vspace{1mm}
\caption{Root causes of embodiment-induced semantic mismatch in embodied AI systems. The table summarizes how semantically correct decisions can systematically fail to ensure physical safety due to structural gaps between language, perception, and physical interaction.}
\label{tab:semantic_mismatch_root_causes}
\end{table*}

\subsubsection{Semantic Correctness in Decision-Making}

\paragraph{Semantic ``Correctness'': Logical Consistency and Intent Matching}
In NLP and dialogue settings, ``correctness'' typically refers to logical consistency and intent matching. Given a user instruction, a semantically correct decision identifies the intended object and goal state and produces a plausible action sequence. Importantly, such evaluation can be performed without modeling physical feasibility, uncertainty, or execution risk. This is appropriate for purely symbolic tasks, but it is incomplete for embodied decision-making.

However, Son et al.~\cite{son2025subtle} (SAFEL) show that semantic correctness is necessary but insufficient for safe embodiment. Even when LLM-generated plans satisfy preconditions and appear causally consistent, execution can still fail in safety-critical ways. Their findings can be summarized as three representative failure types:
\begin{enumerate}
  \item \textbf{Subtle physical risks.} A linguistically reasonable instruction can violate dynamics or contact constraints. For example, ``quickly grasp a fragile glass'' may demand accelerations that exceed friction limits, causing slip or breakage.
  \item \textbf{Context-dependent hazards.} The same action can be safe in one context and unsafe in another. ``Push the object to the table edge'' may be appropriate for cleaning, but becomes dangerous if a person is nearby or if the object could fall.
  \item \textbf{Long-horizon constraint violations.} Plans may satisfy immediate goals while violating intermediate constraints during execution. In assembly, a seemingly valid tightening order can yield poor strength due to unmodeled stress redistribution or deformation effects.
\end{enumerate}
These failures illustrate that semantic validity does not reliably predict physical consequence, especially when safety is determined by geometry, dynamics, and context.

\paragraph{Physical Safety: Constraint Satisfaction and Risk Boundaries}
Physical safety requires real-time satisfaction of multi-dimensional constraints and explicit control of risk. Zhang et al.~\cite{zhang2024safe} organize embodied safety into three pillars:
\begin{enumerate}
  \item \textbf{Kinematic safety:} joint positions, velocities, and accelerations stay within limits and avoid self-collision or singularities.
  \item \textbf{Dynamic safety:} torques and contact forces respect actuation capacity and structural constraints.
  \item \textbf{Task safety:} behavior avoids harm to people, property, and task-specific safety protocols.
\end{enumerate}
These requirements are difficult to express purely in semantic space. Brunke et al.~\cite{brunke2025semantically} propose mapping natural language to formal safety specifications via a semantic safety graph, but such approaches typically rely on expert-defined rule libraries, which limits robustness and generalization in open-domain settings.

\paragraph{Why the Gap Persists: Disembodied Pretraining and Misaligned Metrics}
The semantic--physical gap is reinforced by how LLMs are trained and evaluated. As Xing et al.~\cite{xing2025robust} argue, mainstream corpora contain few causal annotations about physical interaction: verbs such as ``push'' or ``lift'' appear as abstract symbols without contact parameters, frictional regimes, or consequence labels. Even robotics text often describes high-level procedures rather than the low-level control realities that determine safety.

Moreover, standard LLM alignment metrics (helpfulness, harmlessness, honesty) are designed for dialogue and rarely incorporate feedback from the physical world. For example, a response to ``clear the table quickly'' may be judged by fluency and plausibility, not whether it risks breaking objects or injuring people. Liu et al.~\cite{liu2025aligning} report that state-of-the-art models (e.g., GPT-4~\cite{openai2024gpt4technicalreport}, Claude~\cite{Claude4}) degrade substantially on embodied safety benchmarks compared to text-only tasks.
These results show that semantic correctness does not imply safe behavior. We next consider a closely related issue: \emph{even when the intent is correct and the plan is reasonable, identical actions can have different consequences across physical states}. This consequence variability undermines any assumption that a single semantic-to-action mapping is universally safe.

\subsubsection{Consequence Variability of Identical Actions Across Physical States}

\paragraph{State Dependence: Nonlinear State--Action Mapping}
In control theory, system evolution follows $s_{t+1} = f(s_t, a_t)$, where $s_t$ includes pose, velocity, and contact mode, and $a_t$ is the control input. State dependence means the same action $a$ applied at different states $s_i$ and $s_j$ can yield qualitatively different outcomes $f(s_i,a)$ and $f(s_j,a)$, including divergent task success and safety risk. Winter et al.~\cite{winter2024state} identify major sources of such variability in manipulation:
\begin{enumerate}
  \item \textbf{Geometry:} small pose changes shift contact points and normals, changing trajectories under identical forces.
  \item \textbf{Dynamics/contact mode:} different contact states induce different effective dynamics and stability properties.
  \item \textbf{Material properties:} uncertainty in mass, friction, and stiffness necessitates adaptive strategies.
\end{enumerate}
Text-only training does not expose LLMs to this nonlinear phase-space structure. Action words remain discrete symbols without an associated geometry of state space, limiting robust generalization to novel configurations.

\paragraph{Perception Uncertainty and Error Cascades}
Embodied state estimation is imperfect due to occlusion, lighting variation, viewpoint limits, and sensor noise. Jiang et al.~\cite{jiang2025language} report that grounding accuracy drops sharply under partial occlusion in language-guided grasping: a small grounding error can shift the inferred goal state enough to trigger an incorrect plan. Incorrect grounding then propagates downstream, creating cascading errors across planning and control. Prasanna et al.~\cite{prasanna2025perception} argue that perception uncertainty is systemic rather than modular: uncertainty should be propagated to downstream components so agents can detect high-uncertainty states, slow down, re-observe, or adopt conservative behaviors instead of committing to brittle execution.

\paragraph{Cross-Modal Alignment Gaps: Language, Vision, and Action}
Language, vision, and action lie in different representation spaces. Ma et al.~\cite{ma2025survey} note that many Vision-Language-Action (VLA) systems rely on implicit end-to-end alignment and lack explicit consistency constraints. Typical inconsistencies include:
\begin{enumerate}
  \item \textbf{Language--vision mismatch:} open-vocabulary grounding failures; e.g., HiFi-CS improves robustness via open-vocabulary grounding~\cite{bhat2025hifics}.
  \item \textbf{Vision--action mismatch:} visually stable configurations may be dynamically unstable (e.g., near-edge equilibrium), requiring physical prediction; physically grounded VLMs improve but remain error-prone in contact-rich scenes~\cite{gao2024physically}.
  \item \textbf{Language--action mismatch:} vague modifiers cannot be reliably mapped to control parameters, producing overly aggressive or overly conservative execution; task-conditioned action modeling helps but demands labeled data and generalizes imperfectly~\cite{tang2023task}.
\end{enumerate}
These gaps highlight that ``alignment'' must be maintained across representations, not merely within language.

\paragraph{Dynamic Environments: Beyond Quasi-Static Assumptions}
Many VLA pipelines implicitly assume quasi-static scenes during single action executions. In practice, humans and objects move, and the environment can change during execution. Lee et al.~\cite{lee2025dynscene} (DynScene) evaluate robustness under dynamic obstacles, object state changes, and long-horizon drift. Liu et al.~\cite{liu2025roboview} (RoboView-Bias) further show systematic visual bias toward static salient objects, degrading obstacle avoidance and collision prediction in dynamic settings. Together, these results suggest that cross-modal grounding alone is insufficient unless models explicitly account for temporal change and uncertainty.

\begin{tcolorbox}[framingbox]
\textbf{Insight.}
language is discrete and symbolic, whereas physical interaction is continuous and governed by dynamics and constraints. As a result, semantic ``correctness'' ensures intent matching and logical plausibility, but not constraint satisfaction, risk control, or consequence prediction. This gap is further amplified by state (dependent variability—high-dimensional states, perception uncertainty, and dynamic environments) making a single semantic-to-action mapping inherently brittle.
Existing remedies remain partial: accurate physical grounding is hard to obtain, uncertainty propagation incurs significant overhead, cross-modal alignment lacks a unified formulation, and state-dependent learning remains fragile under distribution shift. 
\end{tcolorbox}

\subsection{Action--Consequence Decoupling}

\begin{table*}[t]
\centering
\small
\setlength{\tabcolsep}{6pt}
\begin{tabular}{p{3cm} p{2cm} p{4.8cm} p{3cm} p{2.0cm}}
\toprule
\textbf{Root Cause} &
\textbf{Primary Fracture} &
\textbf{Underlying Mechanism} &
\textbf{Failure Consequence} &
\textbf{Representative Work} \\
\midrule

\textbf{Semantic Gap} &
Validated plans fail in physical execution &
Text-derived knowledge abstracts away forces, contact, friction, and intermediate feedback; language generalizes semantically but not mechanically &
Physically infeasible or unsafe execution despite correct intent &
Roy et al.~\cite{roy2021machine}, LIBERO~\cite{liu2023libero} \\

\addlinespace[2pt]

\textbf{Execution Drift} &
Mismatch between planned and executed behavior &
Planning relies on simplified geometry and quasi-static dynamics; perception error, model mismatch, and environment change accumulate during execution &
Trajectory deviation, delayed reaction, and unsafe control actions &
Pramanick et al.~\cite{pramanick2020decomplex}, Friedrich et al.~\cite{friedrich2022maintenance} \\

\addlinespace[2pt]

\textbf{Risk Accumulation} &
Locally safe actions lead to global hazards &
Multi-level constraints (task, motion, control, interaction) are implicitly coupled; safety is not compositional over time &
Long-horizon safety violations and cascading failures &
He et al.~\cite{he2025provable}, Dandjinou et al.~\cite{dandjinou2025hierarchical} \\

\bottomrule
\end{tabular}
\vspace{1mm}
\caption{Root causes of action--consequence decoupling in embodied AI systems. These causes explain why actions validated as safe or reasonable at decision time may still lead to unsafe physical outcomes during execution.}
\label{tab:action_consequence_root_causes}
\end{table*}

Embodied systems face a cognitive--execution divide: an agent may validate an action as sensible under its internal model, while the physical world ultimately determines whether the resulting consequences are safe. In traditional software, unit tests and formal verification can ensure behavioral correctness because execution follows logical rules. In robotics, however, acting in the world introduces uncertainty, unmodeled contact dynamics, sensing error, and environmental change. Consequently, even an action sequence that passes syntactic checks or high-level safety filters can still produce unsafe outcomes once instantiated in the real environment.

This decoupling is driven by three factors. First, text-derived knowledge transfers poorly to physical interaction, producing a semantic gap. Second, execution drift arises because plans rely on simplified models, while real dynamics and perception are imperfect. Third, multi-level constraints interact over time, enabling risk accumulation: actions that are locally safe can combine into globally hazardous trajectories. Liu et al.~\cite{liu2025aligning} argue that this action--consequence gap is central to embodied safety and must be mitigated by tighter physical modeling and closed-loop feedback.

\subsubsection{Three Root Causes: Fractures from Decision to Execution}

To move beyond surface-level vulnerabilities and isolated failure cases, we now examine the fundamental mechanisms that cause embodied AI systems to fail in practice. While prior sections have separately discussed limitations of LLM reasoning and classical CPS assumptions, these perspectives alone do not fully explain why semantically correct, well-planned actions still lead to unsafe physical behavior. As shown in \autoref{tab:action_consequence_root_causes}, we identify three root causes that systematically undermine embodied autonomy: limited transfer from text to physics, execution drift under simplified models, and long-horizon risk accumulation across interacting constraints.

\paragraph{Root Cause 1: The Semantic Gap---Limited Transfer from Text to Physics}
\textit{Fractured representation from text to mechanics.} Large language models learn from internet text and dialogues, where physical interactions are described abstractly. Roy et al.~\cite{roy2021machine} highlight three deficiencies:
\begin{itemize}
  \item \textbf{Incomplete causal detail:} text rarely encodes contact point, force direction, or friction.
  \item \textbf{Missing intermediate feedback:} descriptions emphasize outcomes over interaction dynamics.
  \item \textbf{Weak OOD coverage:} long-tail objects/materials and rare failure modes are underrepresented.
\end{itemize}

\textit{Transfer bottlenecks in benchmarks.} LIBERO~\cite{liu2023libero} shows success rates drop on novel objects even when tasks are similar, due to implicit assumptions on friction/weight, ambiguous language constraints, and non-stationary dynamics (wear, environment changes). Jaquier et al.~\cite{jaquier2025transfer} similarly argue that true zero-shot physical transfer remains challenging.

\textit{Multimodal integration does not automatically solve transfer.} Cong et al.~\cite{cong2025overview} emphasize multimodal models for embodied intelligence, but fusion introduces alignment inconsistencies. When language suggests ``careful'' but perception indicates fast motion or unstable grasp, the system may fail to resolve conflicts, producing actions that satisfy intent but violate feasibility.

\paragraph{Root Cause 2: Execution Drift---Planning Under Simplified Models}
\textit{Systematic mismatch between planned and executed behavior.} Task and Motion Planning (TAMP) bridges goals to actions but often assumes deterministic geometry and quasi-static dynamics. Drift arises from (i) geometric uncertainty in pose estimates, (ii) dynamic simplification, and (iii) environment changes during execution. While failure-aware planning can include recovery branches, search complexity can become prohibitive in real time.

\textit{Ambiguity amplification in natural instructions.} Complex instructions under-specify timing, priorities, and resource allocation. Decomplex~\cite{pramanick2020decomplex} illustrates how a single instruction can admit serial, parallel, or interleaved plans, each with different safety implications. Friedrich et al.~\cite{friedrich2022maintenance} similarly report that even with prior knowledge, timeliness and missing context can cause execution deviations requiring intervention.

\paragraph{Root Cause 3: Risk Accumulation---Interacting Multi-Level Constraints}
Safety constraints span task, motion, control, and physical interaction. He et al.~\cite{he2025provable} emphasize that constraints are implicitly coupled: satisfying one layer does not guarantee global safety, and sequences of locally safe actions can create global hazards. Hierarchical safety contracts~\cite{dandjinou2025hierarchical} mitigate this via distributed monitoring, but effectiveness depends on contract design and remains challenged in highly dynamic environments.

\textit{Context dependence of safety.} Martinetti et al.~\cite{martinetti2021redefining} note that industrial standards (e.g., ISO 10218, ISO 15066) target structured workcells and only partially address contextual variation in human--robot collaboration. LLM-enabled systems exacerbate this because goals and contexts can change online.

\textit{Long-horizon degradation.} Over extended operation, hardware aging, environment evolution, and knowledge obsolescence degrade safety margins. Quantifying uncertainty growth is possible in restricted settings~\cite{he2025provable}, but remains difficult in open worlds.

\begin{table*}[t]
\centering
\footnotesize
\setlength{\tabcolsep}{6pt}
\renewcommand{\arraystretch}{1.15}
\begin{tabular}{p{4cm} p{3.9cm} p{5.0cm} p{3cm}}
\toprule
\textbf{Failure Mode} &
\textbf{Primary Root Cause(s)} &
\textbf{Mechanism Description} &
\textbf{Representative Work} \\
\midrule

\textbf{Control Failure} 
&
\textbf{Execution Drift} (primary) \newline
Semantic--Physical Gap (secondary)
&
Mismatch between planned trajectories and real actuation limits causes
tracking errors, delayed fault detection, and unsafe motion under saturation
or bandwidth constraints.
&
Braun et al.~\cite{braun2013robots};\,
Ji et al.~\cite{ji2024towards};\,
Nguyen et al.~\cite{nguyen2021robust} \\
\hline

\textbf{Force Feedback Anomalies}
&
\textbf{Semantic--Physical Gap} (primary) \newline
Risk Accumulation (secondary)
&
High-level decisions ignore contact dynamics and uncertainty, leading to
ambiguous force signals, false alarms, and limited generalization of recovery
strategies in contact-rich manipulation.
&
Altan et al.~\cite{altan2021went};\,
Luo et al.~\cite{luo2021endowing};\,
Zhong et al.~\cite{zhong2023detecting} \\
\hline

\textbf{Safety Boundary Breaches}
&
\textbf{Risk Accumulation} (primary)
&
Locally safe actions and delayed detections accumulate across time and layers,
causing fault propagation, cascading failures, and irreversible system-level
safety violations.
&
Xu et al.~\cite{xu2022failure};\,
He et al.~\cite{he2025provable};\,
Martinetti et al.~\cite{martinetti2021redefining} \\
\bottomrule
\end{tabular}
\caption{Mapping between execution-layer failure modes and their underlying root causes in embodied AI systems, with representative empirical and theoretical studies.}
\label{tab:failure_rootcause_mapping}
\end{table*}

\subsubsection{Three Failure Modes: Vulnerabilities at the Execution Layer}

In practice, unsafe behavior rarely stems from a single catastrophic error; instead, it emerges from how decisions interact with control limits, sensing uncertainty, and safety boundaries over time. As shown in \autoref{tab:failure_rootcause_mapping}, we summarize three recurring failure modes that translate upstream mismatches into concrete physical hazards: breakdowns in command-to-motion execution, anomalies in force-feedback and contact handling, and breaches of safety boundaries caused by fault propagation and delayed detection.

\paragraph{Failure Mode 1: Control Failure---From Commands to Motion}
\textit{Tracking failure under actuation limits.} Controllers convert reference trajectories into actuator commands, but saturation and bandwidth limits can induce large deviations. Braun et al.~\cite{braun2013robots} show how demanding commands can exceed compliant actuator capabilities, increasing collision risk. Ji et al.~\cite{ji2024towards} propose failure-aware robust control by modeling actuation constraints as uncertainty sets to preserve safety under worst-case conditions.

\textit{Fault detection latency in redundant systems.} A single joint failure may be recoverable via reconfiguration, but only if detected quickly. Under high-level LLM planning, abnormal signals can be misattributed to legitimate load variations; multi-sensor anomaly detection can reduce latency~\cite{ji2024towards}.

\textit{Robustness versus conservatism.} Robust control trades performance for guarantees. CBF-based methods can preserve safety while maximizing performance~\cite{nguyen2021robust}, but depend on uncertainty bounds. With stochastic LLM-driven decisions, online bound updates help~\cite{ji2024towards}, yet extrapolation failures remain possible.

\paragraph{Failure Mode 2: Force Feedback Anomalies---Breaking the Perception--Execution Loop}
\textit{Anomaly recognition in contact-rich manipulation.} Force/torque signals are noisy and task-dependent. Altan et al.~\cite{altan2021went} model anomalies as time-series classification (e.g., HMMs), but boundary ambiguity induces false alarms that reduce availability.

\textit{Recovery under disturbances.} Luo et al.~\cite{luo2021endowing} separate anomalies into recoverable versus non-recoverable types and design recovery strategies. However, strategies are often task-specific, limiting generalization.

\textit{Collaborative settings.} In HRC, anomalies also arise from human motion. Trajectory-based anomaly detection helps~\cite{zhong2023detecting} but struggles with gradual degradation (e.g., wear), which needs long-horizon statistics.

\paragraph{Failure Mode 3: Safety Boundary Breaches---From Local Safety to System Failure}
\textit{Fault propagation and cascading failures.} System dependability depends on coupling among modules. Xu et al.~\cite{xu2022failure} show that in networked UAV systems, a single fault can propagate via communication to systemic collapse; isolation helps but requires redundancy.
\textit{Residual accumulation and detection delay.} Even good detectors can be too slow under extreme conditions. Combining data-driven and model-based residual methods improves detection but does not eliminate latency risks~\cite{ji2024towards}. 
\textit{Learning prevention skills and OOD failure.} Learning-based prevention from demonstrations can improve common-case robustness~\cite{ak2023learning}, but long-tail failures remain difficult, and rare failures can be the most dangerous.

\begin{tcolorbox}[framingbox]
\textbf{Insight.}
Action--consequence decoupling exposes a persistent gap between what the agent \emph{validates} and what the world \emph{enforces}. Semantic gap, execution drift, and risk accumulation jointly create fault zones from planning to control, where semantically correct actions can still yield hazardous outcomes. Mitigation directions include (i) better physical grounding, (ii) failure-aware planning and recovery, (iii) multi-level monitoring via safety contracts, and (iv) robust control and anomaly detection. However, these approaches are often piecemeal, motivating a system-level view that connects assumptions and guarantees across layers.
\end{tcolorbox}

\subsection{Cross-Layer Misalignment}

Embodied autonomy is rarely implemented as a monolithic, end-to-end policy. Instead, most practical systems adopt a layered architecture: a \emph{semantic layer} (e.g., an LLM or VLA policy) interprets user intent and selects abstract goals, a \emph{planning layer} translates these goals into symbolic plans or continuous trajectories, and a \emph{control layer} executes trajectories under physical constraints. While this decomposition improves modularity and debuggability, it introduces a distinct failure mode that is neither purely a language-model error nor purely a control instability: \emph{cross-layer misalignment}. As shown in \autoref{tab:crosslayer-rootcauses}, the defining feature is that each layer may behave correctly in isolation, yet the composed system behaves unsafely due to incompatible objectives, constraints, or assumptions across layer interfaces~\cite{xu2024llmcps,ghosh2025agenticsafety}.

\begin{table*}[t]
\centering
\caption{Root Causes of Cross-Layer Misalignment in Embodied AI Systems}
\label{tab:crosslayer-rootcauses}
\renewcommand{\arraystretch}{1.15}
\begin{tabular}{p{3.2cm} p{4.3cm} p{5cm} p{3cm}}
\toprule
\textbf{Root Cause} 
& \textbf{Description} 
& \textbf{Why Local Correctness Fails} 
& \textbf{Representative work} \\
\midrule

\textbf{Objective Mismatch} 
& Different layers optimize incompatible objectives (e.g., instruction plausibility vs. efficiency vs. stability). Physical risk is rarely an explicit objective. 
& Each layer reaches a local optimum, but safety is traded off implicitly across interfaces. No layer detects global risk violations. 
& EARBench~\cite{zhu2024earbench}, SafeMind~\cite{chen2025safemind} \\

\textbf{Constraint Incompatibility} 
& Safety constraints are represented differently across layers: soft and contextual at the semantic level, partial and model-dependent at planning, and hard limits at control. 
& Enforcing constraints at a single layer does not guarantee end-to-end safety; violations emerge after abstraction boundaries. 
& Safety Chip~\cite{yang2023plug}, ROBOGUARD~\cite{ravichandran2025roboguard} \\

\textbf{Information Loss at Interfaces} 
& Semantic intent, uncertainty, and safety rationale are compressed into symbolic goals and numeric costs during translation. 
& Downstream layers cannot recover discarded safety context, and upstream layers cannot anticipate physical feasibility or risk. 
& Varley et al.~\cite{varley2024twoarms}, Rais et al.~\cite{rais2024crosslayer} \\

\textbf{Temporal Mismatch} 
& Semantic reasoning, planning, and control operate at different time scales, with asynchronous updates. 
& Symbolic plans become stale as physical states evolve, causing execution under invalid assumptions. 
& Liu et al.~\cite{liu2024longhorizon} \\

\textbf{Distribution Shift and Stochasticity} 
& LLM-driven decision layers introduce stochastic outputs and face distribution shifts unseen during training. 
& Small semantic variations propagate into large execution differences, amplifying cross-layer inconsistency over time. 
& Ghosh et al.~\cite{ghosh2025agenticsafety} \\

\bottomrule
\end{tabular}
\end{table*}

\subsubsection{Intuition: Local Optimality Does Not Compose}

Let the semantic layer produce an abstract goal $g$, the planning layer compute a trajectory $\tau$, and the control layer generate control inputs $u$. Each layer optimizes its own objective:
$$
\begin{aligned}
g^{\ast} &\in \arg\max_{g}\, U_{\mathrm{sem}}(g \mid x),\\
\tau^{\ast} &\in \arg\min_{\tau}\, J_{\mathrm{plan}}(\tau \mid g^{\ast}, \hat{s}),\\
u^{\ast} &\in \arg\min_{u}\, J_{\mathrm{ctrl}}(u \mid \tau^{\ast}, \hat{s}),
\end{aligned}
$$
where $x$ denotes the instruction or context and $\hat{s}$ the estimated system state. Here $U_{\mathrm{sem}}$ is a utility over abstract goals, while $J_{\mathrm{plan}}$ and $J_{\mathrm{ctrl}}$ are cost functionals over trajectories and control inputs, respectively.

Cross-layer misalignment arises when the composed execution
$\mathrm{Execute}(s, g^{\ast}, \tau^{\ast}, u^{\ast})$
violates safety constraints, even though $g^{\ast}$, $\tau^{\ast}$, and $u^{\ast}$ are each locally optimal or feasible under their respective models.
The core issue is that \emph{interfaces are lossy}. Semantic intent, including implicit safety preferences, uncertainty, and contextual nuance, is compressed into symbolic goals. Planning further reduces this information into geometric constraints and scalar costs, and the controller receives only reference trajectories stripped of semantic rationale. Once safety-relevant information is discarded at an interface, downstream layers cannot recover it, and upstream layers cannot anticipate downstream feasibility or risk~\cite{rais2024crosslayer,varley2024twoarms}.

\subsubsection{Sources of Cross-Layer Misalignment}

Cross-layer misalignment is structural. It arises from systematic differences in how layers represent objectives, constraints, and uncertainty.

\paragraph{1) Objective mismatch and risk blind spots}
LLM-level objectives emphasize instruction following and plausibility, planning objectives prioritize numeric efficiency (time, distance, energy), and control objectives emphasize tracking and stability. Physical risk is rarely an explicit optimization target. As a result, systems may trade safety for efficiency without detecting violations locally. Empirical evidence from EARBench shows that LLM-based planners generate semantically coherent plans that embed severe physical hazards~\cite{zhu2024earbench}. SafeMind further demonstrates that safety failures are distributed across semantic understanding, planning, and execution, rather than isolated to a single module~\cite{chen2025safemind}.

\paragraph{2) Constraint incompatibility across layers}
Constraints differ across layers: semantic constraints are soft and contextual; planning constraints are explicit but partial and model-dependent; control constraints are hard physical limits. Enforcing constraints at one interface does not guarantee end-to-end safety. Guardrail approaches such as Safety Chip constrain LLM-generated actions at decision time~\cite{yang2023plug}, while ROBOGUARD grounds high-level safety rules into formal specifications and synthesizes constrained plans~\cite{ravichandran2025roboguard}. However, these approaches assume accurate interfaces and cannot fully account for unmodeled dynamics or execution drift.

\paragraph{3) Information loss in translation}
Language-to-plan translation compresses nuanced intent into numeric heuristics. Plans encode \emph{what} to do but not \emph{why} constraints matter, leaving controllers unable to distinguish benign from safety-critical deviations under disturbance. Modular embodied systems explicitly expose this fragility: when perception shifts or skills behave outside assumed regimes, errors propagate across layers~\cite{varley2024twoarms,ghosh2025agenticsafety}.

\paragraph{4) Temporal mismatch and distribution shift}
Semantic reasoning updates slowly, planning is intermittent, and control operates at high frequency. In dynamic environments, this leads to stale assumptions where symbolic plans remain valid while physical states change. LLM stochasticity further introduces variability across runs. Over long horizons, these effects compound and amplify cross-layer misalignment~\cite{liu2024longhorizon}.

\subsubsection{Manifestations of Cross-Layer Failures}

Empirical studies reveal recurring patterns in which locally correct decisions combine to form unsafe global behavior.

\paragraph{Semantically plausible but physically risky plans}
EARBench~\cite{zhu2024earbench} shows that language-conditioned planners frequently generate plans that satisfy task goals yet involve unstable placements, excessive speeds, or hazardous human proximity. These failures arise from missing risk-aware objectives at the semantic--planning interface rather than from incoherent language output.

\paragraph{Formally safe execution of unsafe intent}
Control-level safety mechanisms (e.g., CBFs) can ensure constraint satisfaction while still executing an unsafe high-level goal. This illustrates a fundamental limitation: control-level safety cannot repair upstream semantic errors~\cite{yu2024cbfinc}.

\paragraph{Locally safe actions with globally unsafe outcomes}
SafeMind~\cite{chen2025safemind} shows that individual actions may satisfy local constraints while long-horizon accumulation leads to global violations. Neither semantic reasoning nor low-level control typically enforces long-term safety invariants.

\begin{tcolorbox}[framingbox]
\textbf{Insight.}
Cross-layer misalignment is a structural failure mode: local correctness at semantic, planning, and control layers does not compose into global safety. Evidence from embodied risk benchmarks and cross-layer analyses shows that safety must be enforced across the autonomy stack. Constraint enforcement, formal guarantees, cross-layer verification, and distributed safety contracts collectively point to a central principle: \emph{robust embodied safety requires explicit cross-layer invariants, shared assumptions, and continuous feedback across the entire pipeline}.

\end{tcolorbox}
% ...


\section{Open Challenges and Countermeasures}
\label{sec:discuss}

\subsection{Safety Protection}

\subsubsection{Inheritance of Biases from Training Data}

Embodied AI systems often inherit biases embedded within their training datasets, which can manifest in discriminatory or unfair behaviors during physical interactions. These biases may stem from imbalances in data representation, such as underrepresentation of certain ethnic groups, genders, or socioeconomic backgrounds in visual or linguistic corpora. In practical deployments, this can result in perceptual errors, such as autonomous vehicles exhibiting higher failure rates in detecting pedestrians from minority demographics, or service robots assigning lower priorities to tasks involving specific user profiles. Such issues not only compromise system reliability but also amplify societal inequities.

\textbf{Challenges:} Identifying and quantifying biases in large-scale, multimodal datasets remains computationally intensive, particularly in real-time embodied scenarios where environmental variability exacerbates biased outcomes. Ensuring fairness across diverse global contexts without degrading overall performance poses a significant hurdle, as debiasing techniques may introduce new inaccuracies.

\textbf{Countermeasures:} To mitigate these risks, strategies include curating diverse and balanced training datasets, employing debiasing algorithms during model fine-tuning, and conducting regular audits using fairness metrics tailored to embodied contexts.

\subsubsection{Hallucination}

Hallucinations in LLMs refer to the generation of factually incorrect or fabricated information, which becomes particularly hazardous in embodied AI where outputs directly influence physical actions. For instance, an agent might hallucinate the presence of an obstacle or misinterpret environmental cues, leading to unsafe navigation or manipulation sequences. This vulnerability is exacerbated in dynamic real-world settings where incomplete or noisy sensory data amplifies the likelihood of erroneous reasoning.

\textbf{Challenges:} Detecting hallucinations in real-time without human intervention is difficult due to the opaque nature of LLM decision-making processes. Balancing hallucination reduction with maintaining creative or adaptive reasoning capabilities presents a trade-off, as overly restrictive safeguards may limit the agent's utility in novel situations.

\textbf{Countermeasures:} Interactive safety emerges as a promising approach, emphasizing real-time human-AI collaboration to detect and correct hallucinations. Techniques involve incorporating feedback loops where human operators validate critical decisions, or using ensemble models to cross-verify outputs before actuation. Further research is needed to integrate these mechanisms seamlessly without imposing excessive latency on time-sensitive operations.

\subsubsection{Multi-Robot Coordination}

Multi-robot systems introduce complexities in coordination, where security flaws can lead to cascading failures across fleets. Vulnerabilities may arise from insecure communication protocols, enabling adversaries to intercept or spoof messages, or from algorithmic conflicts resulting in deadlocks or collisions. In safety-critical applications like warehouse automation or search-and-rescue operations, such issues could cause physical damage or mission failures.

\textbf{Challenges:} Achieving scalable coordination in heterogeneous robot fleets under adversarial conditions is complex, as varying hardware and software configurations complicate standardized security measures. Real-time detection of coordination failures in dynamic environments adds to the difficulty, potentially leading to delayed responses.

\textbf{Countermeasures:} Challenges include ensuring scalable, resilient consensus mechanisms that withstand denial-of-service attacks or Byzantine faults. Addressing these requires robust encryption for inter-robot communications, decentralized identity and state verification protocols, and simulation-based testing to validate coordination under adversarial conditions.

\subsubsection{Human-Factor Risks}

Human interactions with embodied AI systems present unique risks, including over-trust, miscommunication, or intentional misuse. Users may issue vague or contradictory instructions, leading to misinterpretations that result in hazardous actions, or adversaries could exploit social engineering tactics to manipulate agent behavior. Additionally, prolonged reliance on AI can erode human skills, creating dependencies that heighten risks during system failures.

\textbf{Challenges:} Quantifying human-factor risks in diverse user populations is challenging due to variability in user expertise and behavior. Designing systems that adapt to individual user patterns without compromising security introduces privacy concerns and computational overhead.

\textbf{Countermeasures:} Mitigation involves human-centered design principles, such as intuitive interfaces with clear affordances, ambiguity resolution modules in LLMs, and educational programs to foster appropriate user expectations. Ergonomic assessments and user studies are essential to quantify and reduce these human-factor vulnerabilities.

\subsubsection{Cognitive-Perceptual Alignment}

Cognitive-perceptual alignment ensures that an agent's high-level reasoning accurately corresponds to its sensory perceptions, preventing discrepancies that could lead to unsafe executions. Misalignments often occur in multimodal fusion, where linguistic interpretations fail to match visual or tactile inputs, resulting in actions like grasping non-existent objects or ignoring environmental hazards. This challenge is pronounced in unstructured environments with variable lighting or occlusions.

\textbf{Challenges:} Maintaining alignment in noisy or adversarial environments is technically demanding, as perceptual data can be corrupted, leading to cascading errors in reasoning. Developing metrics that accurately measure alignment in real-time without significant computational cost remains an open problem.

\textbf{Countermeasures:} Solutions include advanced alignment techniques, such as contrastive learning to synchronize modalities, and hierarchical architectures that validate perceptual data against cognitive models before action commitment. Ongoing research aims to develop metrics for measuring alignment fidelity in real-time deployments.

\subsection{Investigating Attack/Defense Mechanisms}

\subsubsection{Impossibility of Filters}

The Indistinguishability Challenge posits that boundaries between benign and adversarial inputs are computationally unresolvable. In Embodied AI (EA), the brain-body interface connecting cognitive LLMs to physical actuators cannot be fully secured by external monitors because triggers can be cryptographically hidden in environmental stimuli. Standard defense relies on Cyber-Physical System (CPS) formal verification. However, a Computational Asymmetry exists: creating triggers is polynomial-time, while detection remains intractable.

\textbf{Challenges:} Traditional security wrappers are mathematically insufficient; agents may execute harmful acts while safety filters report no violations. Moving from post-hoc mitigation to intrinsic alignment without degrading the agent's ability to reason in non-deterministic environments. Real-world physical speeds favor attackers. High-frequency control loops cannot afford the overhead of complex auditors, allowing stealth exploits to manifest through micro-actions that appear benign individually. Developing real-time verifiable intelligence that provides computational soundness without compromising reactive motor speed.

\textbf{Countermeasures:} To address these limitations, countermeasures focus on intrinsic model robustness through adversarial training regimes that simulate hidden triggers during pre-deployment phases. Techniques such as certified defenses, leveraging formal methods like abstract interpretation or reachability analysis, can provide probabilistic guarantees against indistinguishable inputs. Additionally, hybrid approaches combining lightweight runtime monitors with offline verification, and exploring neuromorphic computing for faster anomaly detection, offer pathways to balance security and performance. Research into self-healing architectures, where agents autonomously adapt to detected asymmetries, also holds promise for long-term resilience.

\subsubsection{Large Attack Surfaces}

Multimodal systems, including Large Vision-Language Models (LVLMs), expand the attack surface, with vulnerabilities propagating across modalities and limited attack transferability across models requiring broader security assessments.

\textbf{Challenges:} The integration of diverse sensory inputs in embodied AI amplifies the potential entry points for adversaries, making it difficult to achieve comprehensive coverage. Low transferability means defenses must be customized, increasing development costs and complexity, while propagation effects can turn minor perceptual flaws into systemic failures.

\textbf{Countermeasures:} Effective strategies include cross-modal adversarial training to enhance robustness across inputs, standardized benchmarks like extended versions of Harmful-RLbench for multimodal evaluation, and modular defense layers that isolate vulnerabilities. Collaborative industry efforts to share threat intelligence can improve transferability assessments, enabling more generalized protections.

\subsubsection{Irreversible Physical Harm}

Unlike purely digital systems, LLM-driven robots can cause irreversible physical harm. Biases in perception (e.g., via multimodal models like CLIP) or decision-making lead to real-world discrimination, such as lower rescue priorities for certain groups or biased task assignments.

\textbf{Challenges:} The permanence of physical consequences demands defenses that preempt harm, yet distinguishing biased decisions from legitimate ones in real-time remains complex, especially under resource constraints.

\textbf{Countermeasures:} Countermeasures encompass bias-aware training with augmented datasets, real-time ethical oversight modules using rule-based checks, and post-incident forensic tools for accountability. Integrating human rights frameworks into model alignment processes ensures equitable outcomes, while simulation environments allow safe testing of harm scenarios.

\subsubsection{Deciding the Protection Boundary}

The boundary between safety and utility is often misdefined by CPS equivalence, which restricts systems to a fixed set of safe states. True intelligence requires the capacity to occasionally exit these sets to learn from failure.

\textbf{Challenges:} Over-reliance on static safety envelopes may create an evolutionary dead-end, stripping the agent of the friction needed for higher-order reasoning. Designing dynamic security frameworks that permit acceptable risk exploration, shifting the goal from total vulnerability elimination to resilient risk management.

\textbf{Countermeasures:} Adaptive safety systems, such as risk-aware reinforcement learning with dynamic reward functions, enable controlled exploration. Hierarchical governance models, where low-level controllers enforce hard constraints while high-level planners allow flexibility, facilitate this balance. Ethical guidelines from bodies like IEEE promote frameworks that integrate learning from failures without compromising core safety.

\subsection{Policy and Standards}

\subsubsection{Industry-Wide Frameworks}

Shaping policies for ethical AI usage, including UNESCO guidelines, faces hurdles in ensuring proactive, equitable implementation across sectors, particularly for rapidly evolving embodied technologies. Collaborative initiatives, such as those from the Partnership on AI, aim to standardize best practices, but challenges persist in adapting to technological pace.

\textbf{Challenges:} Global inconsistencies in AI policies, such as between the EU AI Act and U.S. approaches, complicate enforcement for embodied systems, with gaps in addressing emerging risks like job displacement or intellectual property in agentic AI.

\textbf{Countermeasures:} Enhanced frameworks should incorporate stakeholder consultations to foster inclusive, sector-specific guidelines that prioritize transparency and accountability.

\subsubsection{Liability and Certification Standards}

Establishing clear liability for physical harms or ethical breaches in embodied AI is challenging, as policies must define certification processes for safety and moral alignment without stifling technological progress. Standards like ISO/TS 15066 for collaborative robots provide baselines, but extensions are needed for LLM-integrated systems.

\textbf{Challenges:} Balancing rigorous certification with innovation incentives is difficult, as overly stringent standards may deter development, while lax ones risk public safety.

\textbf{Countermeasures:} Rigorous certification pipelines, involving third-party audits and simulation-based validations, can clarify liability chains while encouraging innovation through tiered compliance levels.

\subsubsection{Societal Impact}

Policies must tackle broader issues like economic inequality from AI-driven automation and unhealthy dependencies on embodied systems, requiring frameworks that promote inclusive benefits. Initiatives such as universal basic income pilots or reskilling programs address job displacement, while regulations on AI dependency emphasize human oversight.

\textbf{Challenges:} Assessing long-term societal impacts is complex due to unpredictable technological evolution and varying cultural contexts, making policy formulation proactive yet flexible a key difficulty.

\textbf{Countermeasures:} Comprehensive impact assessments, mandated by policies like the EU's AI Act, ensure that deployments consider long-term societal effects, fostering equitable distribution of AI advantages.

\section{Related Work}

Recent years have witnessed growing attention to the \emph{safety and security}
of embodied AI systems, as the tight coupling between intelligent decision-making
and physical execution introduces risks that go beyond traditional AI or CPS
settings. Several surveys have begun to systematically examine these emerging
challenges from complementary perspectives.
Xing et al.~\cite{xing2025towards} present a dedicated survey on embodied AI safety and security, systematically categorizing exogenous and endogenous vulnerabilities, analyzing adversarial attacks on perception, planning, and LLM/LVLM components, and proposing a unified framework and mitigation strategies to enhance robustness and reliability in real-world embodied systems.
Wang et al. ~\cite{wang2025safety} present a comprehensive survey on safety in embodied navigation, covering attack strategies, defense mechanisms, and evaluation methodologies, and outlining open challenges and future research directions for safer and more reliable systems.  
Khalifa et al.~\cite{khalifa2025securing} examine the role of AI in cybersecurity with a focus on embodied AI systems, systematically analyzing their unique security vulnerabilities—from data integrity and privacy to physical attacks—while discussing ethical implications and providing practical guidelines for securing autonomous and embedded AI in real-world environments. Karne et al.~\cite{karne2025embodied} review recent advances in embodied AI for security applications, comparing representative robotic and drone platforms across surveillance, border control, and disaster response, identifying key technical and ethical challenges, and outlining emerging research trends and future directions toward secure-by-design autonomous systems. Perlo et al.~\cite{perlo2025embodied} analyze the risks of embodied AI systems across physical, informational, economic, and social dimensions, evaluate how existing US, EU, and UK policies address these risks, and propose policy recommendations for the safe, accountable, and socially beneficial deployment of embodied AI. 
Huang et al.~\cite{huang2025beyond} present the first holistic security analysis of an embodied AI platform (Unitree Go2), uncovering ten cross-layer vulnerabilities across wireless, software, cloud, and hardware components, and demonstrating that securing embodied intelligence requires system-level defenses beyond LLM alignment.

Beyond safety and security, a separate line of surveys focuses on \emph{embodied
intelligence} more broadly, emphasizing system architectures, learning paradigms,
and evaluation methodologies. 
Sun et al. ~\cite{sun2024comprehensive} provide a comprehensive survey of embodied intelligence, tracing its evolution from philosophical foundations to modern systems, highlighting integrated perception–cognition–behavior advances, key challenges, and future directions centered on large PCB models and the Bcent general agent framework for robust and adaptable embodied systems. Hou et al.~\cite{hou2026survey} propose a systematic evaluation framework for embodied AI that spans the full perception–cognition–planning–action loop, reviewing evaluation targets, platforms, and methodologies, and outlining open challenges toward trustworthy, general, and physically grounded embodied intelligence. In contrast to prior surveys that primarily catalog attacks, defenses, platforms,
or evaluation protocols, our survey takes a \emph{root-cause–oriented and
system-level perspective}. Rather than treating LLM vulnerabilities and CPS
failures as independent problem domains, we analyze how their underlying
assumptions break down when combined in embodied AI systems. We identify
embodiment-induced failure mechanisms—such as semantic–physical mismatch,
action–consequence decoupling, and cross-layer misalignment—that systematically
explain why semantically correct decisions and locally safe behaviors can still
lead to unsafe physical outcomes. 
\section{Conclusion}
\label{sec:conclusion}

This survey argues that the security of embodied AI cannot be fully understood through LLM vulnerabilities or classical CPS failures alone. Instead, many breakdowns arise from embodiment-induced system-level mismatches between semantic reasoning, perception, and physical execution. By synthesizing insights across these domains, we show that safety in embodied intelligence is inherently non-compositional, state-dependent, and vulnerable to error propagation. Securing future embodied AI systems therefore requires a shift from component-level defenses to system-level reasoning about physical risk, uncertainty, and long-horizon behavior.
\noindent

\section*{Acknowledgement}

This research was supported by the Quancheng Laboratory Award (QCL20250106). Any opinions, findings, conclusions, or recommendations expressed are those of the authors and not necessarily of the Quancheng Laboratory. The research is also supported by the Key R\&D Program of Shandong Province (No. 2025CXPT033), and the National Natural Science Foundation of China (No. 62302266, 62232010, U23A20302, U24A20244).

\bibliographystyle{IEEEtranS}
\bibliography{cas-refs}

\end{document}